%% file: paper.tex
\documentstyle[aps,preprint]{revtex}
\begin{document}

\title{Squark-Chargino Production in Polarized Gamma-Proton
Collisions at TeV Energy Scale}
\author{Z. Z. Aydin and O. Yilmaz}
\address{ Ankara University,Faculty of Sciences,Department of
Engineering Physics\\
06100 Tando\u gan,Ankara, Turkey}
\maketitle

\begin{abstract}
The associated production of squarks and charginos in high energy collisions
of the polarized real photons and protons is discussed. We give the cross
sections for different initial beam polarizations and the polarization
asymmetries which can be used to anticipate the masses of squarks and charginos.
\end{abstract}

\section{introduction}

Although the Standard Model (SM) of elementary particles has been successful
with high precision up to the scale of 100 GeV, there are many theoretical
reasons that new physics beyond the SM should exist at TeV scale.
Among the models of new physics, the supersymmetry (SUSY) seems to be one of
the most promising candidates for TeV scale. A number of TeV energy machines
have been proposed or constructing, such as the Large Hadron Collider (LHC),
Next Linear $e^{+}e^{-}$ Collider (NLC) and Linac-Ring type ep and $\gamma$p
machines. The latter one can be realized by using the beam of high energy
photons produced through the Compton backscattering of laser photons off a TeV
energy linear $e^{-}$ (or $e^{+}$) beam. We have already proposed some
Linac-Ring type ep and $\gamma$p machines \cite{Aydin}. Here we concentrate
ourselves only on three of them, i.e., the HERA+ LC, LHC+Linac 1 and LHC+TESLA.
Their calculated center of mass energies and luminosities are given in Table 1.

A supersymmetric SM has a new spectrum of particles called SUSY particles
which are the partners of all the known particles with the spins
differing by $\frac{1}{2}$. Some of the SUSY partners are scalar
leptons (sleptons, $\tilde{l}$), scalar quarks (squarks, $\tilde{q}$),
wino ($\tilde{w}^\pm$), Higgsino ($\tilde{H}^\pm_{1,2}$)
(or mixing of the latter ones, charginos, $\chi^\pm_{1,2}$),
photino ($\tilde{\gamma}$), zino ($\tilde{z}^0$), Higgsino ($\tilde{H}
^0_{1,2}$) (or their mixed states, neutralinos, $\chi^{0}_{1,2,3,4}$ ),
gluino ($\tilde{g}$) etc. It is commonly believed that these SUSY particles
should have masses below 1 TeV. Experiments at the existing colliders have
already put the lower mass limits as $m_{\tilde{q}} > 176$ GeV and
$m_{\chi^\pm}> 99$ GeV.

In this paper we study the associated production of the squarks and
charginos at TeV energy $\gamma$p colliders with polarized beams. We have
already discussed this process with unpolarized beams \cite{Alan}. Many
other processes such as ${\gamma}p\to\tilde{q}\tilde{q}^\ast$X, ${\gamma}p\to
\tilde{q}\tilde{g}$X, ${\gamma}p\to\tilde{q}\tilde{\gamma}( or \tilde{q}
\tilde{z})$X, have already been discussed \cite{buch}.

\section{polarized high energy gamma beam}

A beam of laser photons ($\omega_{0}\approx 1.26$ eV, for example) with high
intensity, about $10^{20}$ photons per pulse, is Compton-backscattered off
high energy electrons ($E_e$=250 GeV, for example) from a linear accelerator
and turns into hard photons with a conversion coefficient close to unity.
The energy of the backscattered photons, $E_{\gamma}$, is restricted by the
kinematic condition $y_{max}= 0.83$ (where $y=E_{\gamma}/E_{e}$) in order
to get rid of background effects, in particular $e^{+}e^{-}$ pair production
in the collision of a laser photon with a backscattered photon in the
conversion region.

The details of the Compton kinematics and calculations of the cross section can
be found in ref \cite{ginz}. The energy spectrum of the high energy
real (backscattered) photons, $f_{{\gamma}/e}(y)$, is given by

\begin{eqnarray}
f_{{\gamma}/e}(y)=\frac{1}{D(\kappa)}\biggr[1-y+\frac{1}{1-y}
-4r(1-r)-\lambda_e\lambda_0 r\kappa(2r-1)(2-y)\biggr]
\end{eqnarray}
where $\kappa=4E_e\omega_0/m_e^2$ and $r=y/\kappa(1-y)$.
Here $\lambda_0$ and $\lambda_e$ are the laser photon and
the electron helicities respectively, and $D(\kappa)$ is

\newpage
\begin{eqnarray}
D(\kappa)&=&\biggr(1-\frac{4}{\kappa}-\frac{8}{\kappa^2}\biggr)
ln(1+\kappa)+\frac{1}{2}+\frac{8}{\kappa}-\frac{1}{2(1+\kappa)^2}\nonumber\\
&+&\lambda_e\lambda_0\biggr[(1+\frac{2}{\kappa})ln(1+\kappa)-
\frac{5}{2}+\frac{1}{1+\kappa}-\frac{1}{2(1+\kappa)^2}\biggr]
\end{eqnarray}
In our numerical calculations, we assume $E_e\omega_0=0.3$ $MeV^2$ or
equivalently $\kappa=4.8$ which corresponds to the optimum value of $
y_{max}=0.83$, as mentioned above.

The energy spectrum, $f_{\gamma /e}(y)$, does essentially depend on the value
$\lambda _e\lambda _0$. In the case of opposite helicities ($\lambda
_e\lambda _0=-1$) the spectrum has a very sharp peak at the high
energy part of the photons.
This allows us to get a highly monochromatic high energy gamma
beam by eliminating low energy part of the spectrum \cite{ginz,borden}.
On the contrary, for the same helicities ($\lambda _e\lambda
_0=+1 $) the spectrum is flat.

The average degree of linear polarization of the photon is proportional to
the degree of linear polarization of the laser. In our calculations, we
assume that the degree of linear polarization of the laser is zero so that
the final photons have only the degree of circular polarization $(\lambda
(y)=<\xi _2>\neq 0$ and $<\xi _1>$$=<\xi _3>=0$ $)$. The circular
polarization of the backscattered photon is given as follows

\begin{eqnarray}
<\xi _2>=\lambda(y)=\frac {(1-2r)(\frac{1}{1-y}+1-y)\lambda_0
+\lambda_e r\kappa \biggr(1+(1-2r)^2(1-y)\biggr)}{\frac{1}{1-y}+1-y-4r(1-r)
-\lambda_0\lambda_e r\kappa(2r-1)(2-y)}
\end{eqnarray}

For the same initial polarizations ($\lambda_0\lambda_e=+1$) , it is seen
that $\lambda(y)\approx +1$, as nearly independent of $y$;
while for the case of the opposite
polarizations ($\lambda_0\lambda_e=-1$),
the curve $\lambda(y)$
smoothly changes from $-1$ to $+1$ as $y$ increases from zero to $0.83$
 \cite{ginz,borden}.

\section{polarized cross-sections for the reaction}

The subprocess contributing to our physical process ${\gamma}p\to
\tilde{w}\tilde{q}$X  is  ${\gamma}q\to\tilde{w}\tilde{q}$.
The invariant amplitude for the specific subprocess ${\gamma}u\to
\tilde{w}^{+}\tilde{d}$ is the sum of the three terms corresponding
to the s-channel $u$ quark exchange, the t-channel $\tilde{w}$ wino
exchange and the u-channel $\tilde{d}$ squark exchange interactions:
\begin{eqnarray}
{\cal M}_a &=&\frac{-iee_q g}{2\hat{s}}\bar{u}(p^{\prime})(1-\gamma_5)
(\not\! p+\not\! k)\not\! \epsilon u(p)\nonumber\\
{\cal M}_b &=&\frac{-iege_{\tilde{w}}}{2(\hat{t}-m^2_{\tilde{w}})}
\bar{u}(p^{\prime})\not\! \epsilon (\not\! p-\not\! k+m_{\tilde{w}})
(1-\gamma_5)u(p)\\
{\cal M}_c &=&\frac{-iee_{\tilde{q}}g}{2(\hat{u}-m^2_{\tilde{q}})}
\bar{u}(p^{\prime})(1-\gamma_5)u(p)(p-p^\prime +k^\prime ).\epsilon\nonumber
\end{eqnarray}
where $e_q$, $e_{\tilde{q}}$ and $e_{\tilde{w}}$ are the quark,
squark and wino charges, and $g=e/sin{\theta_{w}}$ is the weak coupling
constant. Note that we ignore the quark masses.

Since we are interested in the polarized cross-section, we use the following
density matrices for the initial photon and quark:

\begin{eqnarray}
\rho^{(\gamma)}&=&\frac{1}{2}(1+\vec{\xi}.\vec{\sigma})\nonumber\\
\rho^{(q)}&= &u\bar{u} =\not\! p[1+\gamma_{5}(\lambda_{q}+\vec{\xi}_{\perp}.
\vec{\gamma}_{\perp})]
\end{eqnarray}
where ${\xi}_1$, ${\xi}_2$ and ${\xi}_3$ are Stokes parameters. We take into
account only circular polarization for the photon which is defined
by ${\xi}_2$, as has been already mentioned in the previous section.
$\lambda_q$ stands for the helicity of the parton-quark that is
$+1 (-1)$ for the spin directions parallel (anti-parallel) to its
momentum. The last term in the quark density matrix does not contribute,
because after the integration over the azimuthal angle it vanishes.

One can easily obtain the differential cross section for the subprocess
${\gamma}u\to\tilde{w}^+\tilde{d}$  as follows
\begin{eqnarray}
\frac{d\hat{\sigma}}{d\hat{t}}=\frac{\pi \alpha^2}{\hat{s}^2 sin^2
\theta_w}(1+\lambda_{q})\biggr[\frac{d\hat{\sigma_0}}{d\hat{t}}
+\lambda(y)\frac{d\hat{\sigma}_1}{d\hat{t}}\biggr].
\end{eqnarray}
Performing the $d\hat{t}$ integration from $t_{min}$ to $t^{max}$ which
are given by
\begin{eqnarray}
t^{max}_{min}=\frac{1}{2}(m^2_{\tilde{w}}+m^2_{\tilde{q}}-\hat{s})
\biggr[1\mp\sqrt{1-4m^2_{\tilde{w}}m^2_{\tilde{q}}/(m^2_{\tilde{w}}
+m^2_{\tilde{q}}-\hat{s})^2}\biggr]
\end{eqnarray}
we immediately get the total cross section as
\begin{eqnarray}
\hat{\sigma} (m_{\tilde{w}},m_{\tilde{q}},\hat{s},\lambda{(y)})
=\frac{\pi \alpha^2}{\hat{s}^2 sin^2 \theta_w}(1+\lambda_q)
\biggr[\hat{\sigma_{0}}(m_{\tilde{w}},m_{\tilde{q}},\hat{s})
+{\lambda(y)} \hat{\sigma_{1}}(m_{\tilde{w}},m_{\tilde{q}},\hat{s}) \biggr].
\end{eqnarray}
Note that the cross sections ( Eqs(6) and (8)) are zero for $\lambda_{q}
=-1$ because of the fact that we ignore the quark mass. Integrating the
subprocess cross-section $\hat{\sigma}$ over the quark and photon distributions
we obtain the total cross-section for the physical process
$\gamma p\to\tilde{w} \tilde{q}X$ (the new variables are defined by
$\hat{s}\equiv s_{\gamma q}=xys$, $xy=\tau$ and $s\equiv s_{e p}$):

\begin{eqnarray}
\sigma=\int^{0.83}_{(m_{\tilde{w}}+m_{\tilde{q}})^{2}/s} d\tau
\int^{1}_{\tau/0.83} \frac{dx}{x} f_{\gamma /e}(\tau/x) f_{q}(x)
\hat{\sigma}(m_{\tilde{w}},m_{\tilde{q}},\hat{s},\lambda(\tau/x))
\end{eqnarray}
where the photon distribution function, $f_{{\gamma}/e}(y)$, is actually the
normalized differential cross-section of the Compton backscattering, Eq.(1)
; $f_q (x)$ is the distribution of quarks inside the proton. We set
$\lambda_q=+1$ and $f_q (x)\to u^{+} (x)=\frac{1}{2}(u_{unp}+{\Delta}u_{pol})$
for the $u$-type valence quark distribution. In our numerical calculations,
we use the distribution functions given in Ref. \cite{alta}
and Ref. \cite{kell} for the unpolarized and polarized up-quarks, respectively:
\begin{eqnarray}
u_{unp}(x)= 2.751 x^{-0.412}(1-x)^{2.69}
\end{eqnarray}
\begin{eqnarray}
\Delta u_{pol}(x)= 2.132 x^{-0.2}(1-x)^{2.40}
\end{eqnarray}
Performing the integrations in Eq.(9) numerically we obtain
the total cross-section for the associated wino-squark production.
We plot the dependence of the total cross-sections on the masses of
the SUSY particles for various proposed $\gamma p$ colliders in
Figs. 1(a-c) for $\lambda_0 \lambda_e =+1$ and $-1$.
By taking 100 events per running year
as observation limit for a SUSY particle, one can easily find the upper
discovery mass limits from these figures using the luminosities of
the proposed $\gamma p$ colliders given in Table 1. These discovery
limits are tabulated in the same table.

\section{asymmetry}
It may be more interesting to use a polarization asymmetry in determining
the masses of SUSY particles. Such an asymmetry can be defined with
respect to the product of the polarizations of the laser photon and
the electron as follows
\begin{eqnarray}
A=\frac{\sigma_{-}-\sigma_{+}}{\sigma_{-}+\sigma_{+}}
\end{eqnarray}
where $\sigma_{+}$ and $\sigma_{-}$ are the polarized total cross-sections
given in Figs. 1(a-c).
The results of the polarization asymmetry are shown in Figs. 2(a-c) for
three colliders.

\section{conclusion}
If one compares the curves $\sigma_{+}$ and $\sigma_{-}$ in Figs. 1(a-c)
for each collider one sees that the
polarized cross-sections for different polarization do not differ much
from each other and also from the unpolarized ones. Therefore, the discovery
mass limits for SUSY partners obtained with polarized beams are nearly
equal to those obtained with unpolarized beams. But the polarization
asymmetry is highly sensitive to the wino
and the squark masses and as high as 0.4 for all cases.
Especially in the case of $m_{\tilde{q}}=250$ $GeV$,
the asymmetry parameter A is around 0.6 for the higher wino masses.

The signature of the associated $\tilde{w}^{+} \tilde{d}$ production
will depend on the mass spectrum of SUSY particles. It is generally assumed
that the photino and sneutrino are the lightest SUSY particles and
that the hierarchy of the squark masses is similar to that of quarks.
With these assumption we have the following decays for the case
$m_{\tilde{q}}=m_{\tilde{w}}$:

$\tilde{d}\to d\tilde{\gamma}$, $d\tilde{g}$   and
  $\tilde{w}\to l^{+}\tilde{\nu}$,  $\nu\tilde{l}^{+}$,  $W^{+}\tilde{\gamma}$

By taking into account the further decays $\tilde{l}^{+}\to l\tilde{\gamma}$
and $W^{+}\to l^{+}\nu, q \bar{q}$ we arrive at the ultimate final states
as

$l^{+}+n$ $jets(n=1,3,5)+$ large missing energy and missing $P_{T}$

The main background for the final state $l^{+}+jet+P_{T}^{miss}$ will come
from the process $\gamma q\to W q \to l^{+}\nu q$; but this background
may be reduced, in principle,by the cut $P_{T}^{miss}>45$ $GeV$ if
$m_{\tilde{w}} \gg m_{W}$.

\figure{FIG.1.(a-c) Squark-wino production cross-sections as
a function of the mass. Two thin curves on the left
stand for the collider HERA+LC, middle curves for
LHC+Linac 1 and two curves on the right for LHC+TESLA.
A little bit higher (lower) curve of each twin is for
$\lambda_{0} \lambda_{e}=-1$
($\lambda_{0} \lambda_{e}=+1$), i.e., $\sigma_{-}$ ($\sigma_{+}$)}.

\figure{FIG.2.(a-c) Polarization asymmetries as a function of the mass.
Thin curve stands for the collider HERA+LC, middle curve for LHC+Linac 1
and thick curve for LHC+TESLA}.

\begin{table}
\caption{Energy and luminosity values of different $\gamma$p
colliders. The discovery mass limits
for squarks and winos are given in last three columns for
$\lambda_0 \lambda_e=+1$ ($\lambda_0 \lambda_e=-1$).}

\begin{tabular}{|l|l|l|l|l|l|}
\hline
{\bf Machines} &
\begin{tabular}{l}
$\sqrt{s_{ep}}$ \\ 
$(TeV)$%
\end{tabular}
&
\begin{tabular}{l}
${\cal L}_{\gamma p}$ \\
$(10^{30} cm^{-2}s^{-1})$%
\end{tabular}
&
\begin{tabular}{l}
$m_{\tilde{q}}=m_{\tilde{w}}$ \\
$(TeV)$%
\end{tabular}
& 
\begin{tabular}{l}
$m_{\tilde{q}}=0.25$ \\
$m_{\tilde{w}}\quad (TeV)$%
\end{tabular}
& 
\begin{tabular}{l}
$m_{\tilde{w}}=0.10$ \\
$m_{\tilde{q}}\quad (TeV)$%
\end{tabular}
\\ \hline
{\bf HERA+LC} & $1.28$ & $25$ &$0.23(0.25)$ & $0.18(0.23)$ & $0.35(0.40)$ \\ \hline
{\bf LHC+Linac 1} & $3.04$  & $500$ & $0.65(0.70)$ & $1.00(1.15)$ & $1.33(1.48)$ \\ \hline
{\bf LHC+TESLA} & $5.55$ & $500$ & $0.95(1.05)$ & $1.55(1.70)$ & $2.15(2.50)$ \\ \hline
\end{tabular}
\end{table}

\begin{center}
\input{fig1a}
\smallskip

Fig.1.a

\input{fig1b}
\smallskip

Fig.1.b

\input{fig1c}
\smallskip

Fig.1.c

\input{fig3a}
\smallskip

Fig.2.a

\input{fig3b}
\smallskip

Fig.2.b

\input{fig3c}
\smallskip

Fig.2.c

\end{center}
\end{document}

%% file: fig3a.tex
% GNUPLOT: LaTeX picture
\setlength{\unitlength}{0.240900pt}
\begin{picture}(1500,900)(0,0)
\tenrm
\ifx\plotpoint\undefined\newsavebox{\plotpoint}\fi
\put(264,473){\line(1,0){1172}}
\put(264,113){\line(1,0){20}}
\put(1436,113){\line(-1,0){20}}
\put(242,113){\makebox(0,0)[r]{-0.6}}
\put(264,233){\line(1,0){20}}
\put(1436,233){\line(-1,0){20}}
\put(242,233){\makebox(0,0)[r]{-0.4}}
\put(264,353){\line(1,0){20}}
\put(1436,353){\line(-1,0){20}}
\put(242,353){\makebox(0,0)[r]{-0.2}}
\put(264,473){\line(1,0){20}}
\put(1436,473){\line(-1,0){20}}
\put(242,473){\makebox(0,0)[r]{0}}
\put(264,592){\line(1,0){20}}
\put(1436,592){\line(-1,0){20}}
\put(242,592){\makebox(0,0)[r]{0.2}}
\put(264,712){\line(1,0){20}}
\put(1436,712){\line(-1,0){20}}
\put(242,712){\makebox(0,0)[r]{0.4}}
\put(264,832){\line(1,0){20}}
\put(1436,832){\line(-1,0){20}}
\put(242,832){\makebox(0,0)[r]{0.6}}
\put(264,113){\line(0,1){20}}
\put(264,832){\line(0,-1){20}}
\put(264,68){\makebox(0,0){50}}
\put(408,113){\line(0,1){20}}
\put(408,832){\line(0,-1){20}}
\put(408,68){\makebox(0,0){350}}
\put(551,113){\line(0,1){20}}
\put(551,832){\line(0,-1){20}}
\put(551,68){\makebox(0,0){650}}
\put(695,113){\line(0,1){20}}
\put(695,832){\line(0,-1){20}}
\put(695,68){\makebox(0,0){950}}
\put(838,113){\line(0,1){20}}
\put(838,832){\line(0,-1){20}}
\put(838,68){\makebox(0,0){1250}}
\put(982,113){\line(0,1){20}}
\put(982,832){\line(0,-1){20}}
\put(982,68){\makebox(0,0){1550}}
\put(1125,113){\line(0,1){20}}
\put(1125,832){\line(0,-1){20}}
\put(1125,68){\makebox(0,0){1850}}
\put(1269,113){\line(0,1){20}}
\put(1269,832){\line(0,-1){20}}
\put(1269,68){\makebox(0,0){2150}}
\put(1412,113){\line(0,1){20}}
\put(1412,832){\line(0,-1){20}}
\put(1412,68){\makebox(0,0){2450}}
\put(264,113){\line(1,0){1172}}
\put(1436,113){\line(0,1){719}}
\put(1436,832){\line(-1,0){1172}}
\put(264,832){\line(0,-1){719}}
\put(45,472){\makebox(0,0)[l]{\shortstack{$A(m_{\tilde{q}})$}}}
\put(850,23){\makebox(0,0){$m_{\tilde{q}}=m_{\tilde{w}}$}}
\sbox{\plotpoint}{\rule[-0.200pt]{0.400pt}{0.400pt}}%
\put(264,504){\usebox{\plotpoint}}
\put(264,504){\rule[-0.200pt]{0.400pt}{0.642pt}}
\put(265,506){\rule[-0.200pt]{0.400pt}{0.642pt}}
\put(266,509){\rule[-0.200pt]{0.400pt}{0.642pt}}
\put(267,511){\rule[-0.200pt]{0.400pt}{0.642pt}}
\put(268,514){\rule[-0.200pt]{0.400pt}{0.642pt}}
\put(269,517){\rule[-0.200pt]{0.400pt}{0.642pt}}
\put(270,520){\rule[-0.200pt]{0.400pt}{0.642pt}}
\put(271,522){\rule[-0.200pt]{0.400pt}{0.642pt}}
\put(272,525){\rule[-0.200pt]{0.400pt}{0.642pt}}
\put(273,528){\rule[-0.200pt]{0.400pt}{0.642pt}}
\put(274,530){\rule[-0.200pt]{0.400pt}{0.642pt}}
\put(275,533){\rule[-0.200pt]{0.400pt}{0.642pt}}
\put(276,536){\rule[-0.200pt]{0.400pt}{0.522pt}}
\put(277,538){\rule[-0.200pt]{0.400pt}{0.522pt}}
\put(278,540){\rule[-0.200pt]{0.400pt}{0.522pt}}
\put(279,542){\rule[-0.200pt]{0.400pt}{0.522pt}}
\put(280,544){\rule[-0.200pt]{0.400pt}{0.522pt}}
\put(281,546){\rule[-0.200pt]{0.400pt}{0.522pt}}
\put(282,549){\rule[-0.200pt]{0.400pt}{0.522pt}}
\put(283,551){\rule[-0.200pt]{0.400pt}{0.522pt}}
\put(284,553){\rule[-0.200pt]{0.400pt}{0.522pt}}
\put(285,555){\rule[-0.200pt]{0.400pt}{0.522pt}}
\put(286,557){\rule[-0.200pt]{0.400pt}{0.522pt}}
\put(287,559){\rule[-0.200pt]{0.400pt}{0.522pt}}
\put(288,562){\rule[-0.200pt]{0.400pt}{0.442pt}}
\put(289,563){\rule[-0.200pt]{0.400pt}{0.442pt}}
\put(290,565){\rule[-0.200pt]{0.400pt}{0.442pt}}
\put(291,567){\rule[-0.200pt]{0.400pt}{0.442pt}}
\put(292,569){\rule[-0.200pt]{0.400pt}{0.442pt}}
\put(293,571){\rule[-0.200pt]{0.400pt}{0.442pt}}
\put(294,572){\rule[-0.200pt]{0.400pt}{0.442pt}}
\put(295,574){\rule[-0.200pt]{0.400pt}{0.442pt}}
\put(296,576){\rule[-0.200pt]{0.400pt}{0.442pt}}
\put(297,578){\rule[-0.200pt]{0.400pt}{0.442pt}}
\put(298,580){\rule[-0.200pt]{0.400pt}{0.442pt}}
\put(299,582){\rule[-0.200pt]{0.400pt}{0.442pt}}
\put(300,583){\usebox{\plotpoint}}
\put(301,585){\usebox{\plotpoint}}
\put(302,587){\usebox{\plotpoint}}
\put(303,588){\usebox{\plotpoint}}
\put(304,590){\usebox{\plotpoint}}
\put(305,591){\usebox{\plotpoint}}
\put(306,593){\usebox{\plotpoint}}
\put(307,595){\usebox{\plotpoint}}
\put(308,596){\usebox{\plotpoint}}
\put(309,598){\usebox{\plotpoint}}
\put(310,599){\usebox{\plotpoint}}
\put(311,601){\usebox{\plotpoint}}
\put(312,602){\usebox{\plotpoint}}
\put(313,604){\usebox{\plotpoint}}
\put(314,605){\usebox{\plotpoint}}
\put(315,607){\usebox{\plotpoint}}
\put(316,608){\usebox{\plotpoint}}
\put(317,610){\usebox{\plotpoint}}
\put(318,611){\usebox{\plotpoint}}
\put(319,612){\usebox{\plotpoint}}
\put(320,614){\usebox{\plotpoint}}
\put(321,615){\usebox{\plotpoint}}
\put(322,617){\usebox{\plotpoint}}
\put(323,618){\usebox{\plotpoint}}
\put(324,620){\usebox{\plotpoint}}
\put(325,621){\usebox{\plotpoint}}
\put(326,622){\usebox{\plotpoint}}
\put(327,623){\usebox{\plotpoint}}
\put(328,625){\usebox{\plotpoint}}
\put(329,626){\usebox{\plotpoint}}
\put(330,627){\usebox{\plotpoint}}
\put(331,628){\usebox{\plotpoint}}
\put(332,630){\usebox{\plotpoint}}
\put(333,631){\usebox{\plotpoint}}
\put(334,632){\usebox{\plotpoint}}
\put(335,633){\usebox{\plotpoint}}
\put(336,635){\usebox{\plotpoint}}
\put(337,636){\usebox{\plotpoint}}
\put(338,637){\usebox{\plotpoint}}
\put(339,638){\usebox{\plotpoint}}
\put(340,639){\usebox{\plotpoint}}
\put(341,640){\usebox{\plotpoint}}
\put(342,642){\usebox{\plotpoint}}
\put(343,643){\usebox{\plotpoint}}
\put(344,644){\usebox{\plotpoint}}
\put(345,645){\usebox{\plotpoint}}
\put(346,646){\usebox{\plotpoint}}
\put(347,647){\usebox{\plotpoint}}
\put(348,649){\usebox{\plotpoint}}
\put(349,650){\usebox{\plotpoint}}
\put(350,651){\usebox{\plotpoint}}
\put(351,652){\usebox{\plotpoint}}
\put(352,653){\usebox{\plotpoint}}
\put(353,654){\usebox{\plotpoint}}
\put(354,655){\usebox{\plotpoint}}
\put(355,656){\usebox{\plotpoint}}
\put(356,657){\usebox{\plotpoint}}
\put(357,658){\usebox{\plotpoint}}
\put(358,659){\usebox{\plotpoint}}
\put(359,660){\usebox{\plotpoint}}
\put(360,661){\usebox{\plotpoint}}
\put(361,662){\usebox{\plotpoint}}
\put(362,663){\usebox{\plotpoint}}
\put(363,664){\usebox{\plotpoint}}
\put(364,665){\usebox{\plotpoint}}
\put(365,666){\usebox{\plotpoint}}
\put(366,667){\usebox{\plotpoint}}
\put(367,668){\usebox{\plotpoint}}
\put(368,669){\usebox{\plotpoint}}
\put(369,670){\usebox{\plotpoint}}
\put(370,671){\usebox{\plotpoint}}
\put(372,672){\usebox{\plotpoint}}
\put(373,673){\usebox{\plotpoint}}
\put(374,674){\usebox{\plotpoint}}
\put(376,675){\usebox{\plotpoint}}
\put(377,676){\usebox{\plotpoint}}
\put(378,677){\usebox{\plotpoint}}
\put(380,678){\usebox{\plotpoint}}
\put(381,679){\usebox{\plotpoint}}
\put(382,680){\usebox{\plotpoint}}
\put(384,681){\rule[-0.200pt]{0.413pt}{0.400pt}}
\put(385,682){\rule[-0.200pt]{0.413pt}{0.400pt}}
\put(387,683){\rule[-0.200pt]{0.413pt}{0.400pt}}
\put(389,684){\rule[-0.200pt]{0.413pt}{0.400pt}}
\put(390,685){\rule[-0.200pt]{0.413pt}{0.400pt}}
\put(392,686){\rule[-0.200pt]{0.413pt}{0.400pt}}
\put(394,687){\rule[-0.200pt]{0.413pt}{0.400pt}}
\put(396,688){\rule[-0.200pt]{0.578pt}{0.400pt}}
\put(398,689){\rule[-0.200pt]{0.578pt}{0.400pt}}
\put(400,690){\rule[-0.200pt]{0.578pt}{0.400pt}}
\put(403,691){\rule[-0.200pt]{0.578pt}{0.400pt}}
\put(405,692){\rule[-0.200pt]{0.578pt}{0.400pt}}
\put(407,693){\rule[-0.200pt]{2.650pt}{0.400pt}}
\put(419,694){\rule[-0.200pt]{0.964pt}{0.400pt}}
\put(423,693){\rule[-0.200pt]{0.964pt}{0.400pt}}
\put(427,692){\rule[-0.200pt]{0.964pt}{0.400pt}}
\put(431,691){\usebox{\plotpoint}}
\put(432,690){\usebox{\plotpoint}}
\put(434,689){\usebox{\plotpoint}}
\put(435,688){\usebox{\plotpoint}}
\put(437,687){\usebox{\plotpoint}}
\put(438,686){\usebox{\plotpoint}}
\put(440,685){\usebox{\plotpoint}}
\put(441,684){\usebox{\plotpoint}}
\put(443,681){\usebox{\plotpoint}}
\put(444,680){\usebox{\plotpoint}}
\put(445,678){\usebox{\plotpoint}}
\put(446,677){\usebox{\plotpoint}}
\put(447,675){\usebox{\plotpoint}}
\put(448,674){\usebox{\plotpoint}}
\put(449,673){\usebox{\plotpoint}}
\put(450,671){\usebox{\plotpoint}}
\put(451,670){\usebox{\plotpoint}}
\put(452,668){\usebox{\plotpoint}}
\put(453,667){\usebox{\plotpoint}}
\put(454,666){\usebox{\plotpoint}}
\put(455,663){\rule[-0.200pt]{0.400pt}{0.602pt}}
\put(456,661){\rule[-0.200pt]{0.400pt}{0.602pt}}
\put(457,658){\rule[-0.200pt]{0.400pt}{0.602pt}}
\put(458,656){\rule[-0.200pt]{0.400pt}{0.602pt}}
\put(459,653){\rule[-0.200pt]{0.400pt}{0.602pt}}
\put(460,651){\rule[-0.200pt]{0.400pt}{0.602pt}}
\put(461,648){\rule[-0.200pt]{0.400pt}{0.602pt}}
\put(462,646){\rule[-0.200pt]{0.400pt}{0.602pt}}
\put(463,643){\rule[-0.200pt]{0.400pt}{0.602pt}}
\put(464,641){\rule[-0.200pt]{0.400pt}{0.602pt}}
\put(465,638){\rule[-0.200pt]{0.400pt}{0.602pt}}
\put(466,636){\rule[-0.200pt]{0.400pt}{0.602pt}}
\put(467,631){\rule[-0.200pt]{0.400pt}{1.024pt}}
\put(468,627){\rule[-0.200pt]{0.400pt}{1.024pt}}
\put(469,623){\rule[-0.200pt]{0.400pt}{1.024pt}}
\put(470,619){\rule[-0.200pt]{0.400pt}{1.024pt}}
\put(471,614){\rule[-0.200pt]{0.400pt}{1.024pt}}
\put(472,610){\rule[-0.200pt]{0.400pt}{1.024pt}}
\put(473,606){\rule[-0.200pt]{0.400pt}{1.024pt}}
\put(474,602){\rule[-0.200pt]{0.400pt}{1.024pt}}
\put(475,597){\rule[-0.200pt]{0.400pt}{1.024pt}}
\put(476,593){\rule[-0.200pt]{0.400pt}{1.024pt}}
\put(477,589){\rule[-0.200pt]{0.400pt}{1.024pt}}
\put(478,585){\rule[-0.200pt]{0.400pt}{1.024pt}}
\put(479,577){\rule[-0.200pt]{0.400pt}{1.726pt}}
\put(480,570){\rule[-0.200pt]{0.400pt}{1.726pt}}
\put(481,563){\rule[-0.200pt]{0.400pt}{1.726pt}}
\put(482,556){\rule[-0.200pt]{0.400pt}{1.726pt}}
\put(483,549){\rule[-0.200pt]{0.400pt}{1.726pt}}
\put(484,541){\rule[-0.200pt]{0.400pt}{1.726pt}}
\put(485,534){\rule[-0.200pt]{0.400pt}{1.726pt}}
\put(486,527){\rule[-0.200pt]{0.400pt}{1.726pt}}
\put(487,520){\rule[-0.200pt]{0.400pt}{1.726pt}}
\put(488,513){\rule[-0.200pt]{0.400pt}{1.726pt}}
\put(489,506){\rule[-0.200pt]{0.400pt}{1.726pt}}
\put(490,499){\rule[-0.200pt]{0.400pt}{1.726pt}}
\put(491,486){\rule[-0.200pt]{0.400pt}{3.071pt}}
\put(492,473){\rule[-0.200pt]{0.400pt}{3.071pt}}
\put(493,460){\rule[-0.200pt]{0.400pt}{3.071pt}}
\put(494,448){\rule[-0.200pt]{0.400pt}{3.071pt}}
\put(495,435){\rule[-0.200pt]{0.400pt}{3.071pt}}
\put(496,422){\rule[-0.200pt]{0.400pt}{3.071pt}}
\put(497,409){\rule[-0.200pt]{0.400pt}{3.071pt}}
\put(498,397){\rule[-0.200pt]{0.400pt}{3.071pt}}
\put(499,384){\rule[-0.200pt]{0.400pt}{3.071pt}}
\put(500,371){\rule[-0.200pt]{0.400pt}{3.071pt}}
\put(501,358){\rule[-0.200pt]{0.400pt}{3.071pt}}
\put(502,346){\rule[-0.200pt]{0.400pt}{3.071pt}}
\put(503,320){\rule[-0.200pt]{0.400pt}{6.237pt}}
\put(504,294){\rule[-0.200pt]{0.400pt}{6.237pt}}
\put(505,268){\rule[-0.200pt]{0.400pt}{6.237pt}}
\put(506,242){\rule[-0.200pt]{0.400pt}{6.237pt}}
\put(507,216){\rule[-0.200pt]{0.400pt}{6.237pt}}
\put(508,190){\rule[-0.200pt]{0.400pt}{6.237pt}}
\put(509,164){\rule[-0.200pt]{0.400pt}{6.237pt}}
\put(510,138){\rule[-0.200pt]{0.400pt}{6.237pt}}
\put(511,113){\rule[-0.200pt]{0.400pt}{6.237pt}}
\put(512,113){\usebox{\plotpoint}}
\sbox{\plotpoint}{\rule[-0.500pt]{1.000pt}{1.000pt}}%
\put(264,439){\usebox{\plotpoint}}
\put(264,439){\usebox{\plotpoint}}
\put(265,441){\usebox{\plotpoint}}
\put(266,443){\usebox{\plotpoint}}
\put(267,445){\usebox{\plotpoint}}
\put(268,447){\usebox{\plotpoint}}
\put(269,449){\usebox{\plotpoint}}
\put(270,451){\usebox{\plotpoint}}
\put(271,454){\usebox{\plotpoint}}
\put(272,456){\usebox{\plotpoint}}
\put(273,458){\usebox{\plotpoint}}
\put(274,460){\usebox{\plotpoint}}
\put(275,462){\usebox{\plotpoint}}
\put(276,464){\usebox{\plotpoint}}
\put(277,467){\usebox{\plotpoint}}
\put(278,469){\usebox{\plotpoint}}
\put(279,471){\usebox{\plotpoint}}
\put(280,473){\usebox{\plotpoint}}
\put(281,475){\usebox{\plotpoint}}
\put(282,477){\usebox{\plotpoint}}
\put(283,480){\usebox{\plotpoint}}
\put(284,482){\usebox{\plotpoint}}
\put(285,484){\usebox{\plotpoint}}
\put(286,486){\usebox{\plotpoint}}
\put(287,488){\usebox{\plotpoint}}
\put(288,490){\usebox{\plotpoint}}
\put(289,492){\usebox{\plotpoint}}
\put(290,493){\usebox{\plotpoint}}
\put(291,494){\usebox{\plotpoint}}
\put(292,496){\usebox{\plotpoint}}
\put(293,497){\usebox{\plotpoint}}
\put(294,498){\usebox{\plotpoint}}
\put(295,500){\usebox{\plotpoint}}
\put(296,501){\usebox{\plotpoint}}
\put(297,502){\usebox{\plotpoint}}
\put(298,503){\usebox{\plotpoint}}
\put(299,505){\usebox{\plotpoint}}
\put(300,506){\usebox{\plotpoint}}
\put(301,507){\usebox{\plotpoint}}
\put(302,509){\usebox{\plotpoint}}
\put(303,510){\usebox{\plotpoint}}
\put(304,511){\usebox{\plotpoint}}
\put(305,512){\usebox{\plotpoint}}
\put(306,514){\usebox{\plotpoint}}
\put(307,515){\usebox{\plotpoint}}
\put(308,516){\usebox{\plotpoint}}
\put(309,518){\usebox{\plotpoint}}
\put(310,519){\usebox{\plotpoint}}
\put(311,520){\usebox{\plotpoint}}
\put(312,522){\usebox{\plotpoint}}
\put(313,523){\usebox{\plotpoint}}
\put(314,524){\usebox{\plotpoint}}
\put(315,525){\usebox{\plotpoint}}
\put(316,526){\usebox{\plotpoint}}
\put(317,527){\usebox{\plotpoint}}
\put(318,528){\usebox{\plotpoint}}
\put(319,529){\usebox{\plotpoint}}
\put(320,530){\usebox{\plotpoint}}
\put(321,531){\usebox{\plotpoint}}
\put(322,532){\usebox{\plotpoint}}
\put(323,533){\usebox{\plotpoint}}
\put(324,534){\usebox{\plotpoint}}
\put(325,535){\usebox{\plotpoint}}
\put(326,536){\usebox{\plotpoint}}
\put(327,537){\usebox{\plotpoint}}
\put(328,538){\usebox{\plotpoint}}
\put(329,539){\usebox{\plotpoint}}
\put(330,540){\usebox{\plotpoint}}
\put(331,541){\usebox{\plotpoint}}
\put(332,542){\usebox{\plotpoint}}
\put(333,543){\usebox{\plotpoint}}
\put(334,544){\usebox{\plotpoint}}
\put(335,545){\usebox{\plotpoint}}
\put(336,546){\usebox{\plotpoint}}
\put(337,547){\usebox{\plotpoint}}
\put(338,548){\usebox{\plotpoint}}
\put(339,549){\usebox{\plotpoint}}
\put(340,550){\usebox{\plotpoint}}
\put(341,551){\usebox{\plotpoint}}
\put(342,552){\usebox{\plotpoint}}
\put(343,553){\usebox{\plotpoint}}
\put(345,554){\usebox{\plotpoint}}
\put(346,555){\usebox{\plotpoint}}
\put(347,556){\usebox{\plotpoint}}
\put(348,557){\usebox{\plotpoint}}
\put(349,558){\usebox{\plotpoint}}
\put(350,559){\usebox{\plotpoint}}
\put(351,560){\usebox{\plotpoint}}
\put(353,561){\usebox{\plotpoint}}
\put(354,562){\usebox{\plotpoint}}
\put(355,563){\usebox{\plotpoint}}
\put(356,564){\usebox{\plotpoint}}
\put(357,565){\usebox{\plotpoint}}
\put(358,566){\usebox{\plotpoint}}
\put(359,567){\usebox{\plotpoint}}
\put(361,568){\usebox{\plotpoint}}
\put(362,569){\usebox{\plotpoint}}
\put(364,570){\usebox{\plotpoint}}
\put(365,571){\usebox{\plotpoint}}
\put(366,572){\usebox{\plotpoint}}
\put(368,573){\usebox{\plotpoint}}
\put(369,574){\usebox{\plotpoint}}
\put(370,575){\usebox{\plotpoint}}
\put(372,576){\usebox{\plotpoint}}
\put(373,577){\usebox{\plotpoint}}
\put(374,578){\usebox{\plotpoint}}
\put(376,579){\usebox{\plotpoint}}
\put(377,580){\usebox{\plotpoint}}
\put(378,581){\usebox{\plotpoint}}
\put(380,582){\usebox{\plotpoint}}
\put(381,583){\usebox{\plotpoint}}
\put(382,584){\usebox{\plotpoint}}
\put(384,585){\usebox{\plotpoint}}
\put(385,586){\usebox{\plotpoint}}
\put(387,587){\usebox{\plotpoint}}
\put(388,588){\usebox{\plotpoint}}
\put(390,589){\usebox{\plotpoint}}
\put(391,590){\usebox{\plotpoint}}
\put(393,591){\usebox{\plotpoint}}
\put(394,592){\usebox{\plotpoint}}
\put(396,593){\usebox{\plotpoint}}
\put(397,594){\usebox{\plotpoint}}
\put(399,595){\usebox{\plotpoint}}
\put(400,596){\usebox{\plotpoint}}
\put(402,597){\usebox{\plotpoint}}
\put(403,598){\usebox{\plotpoint}}
\put(405,599){\usebox{\plotpoint}}
\put(406,600){\usebox{\plotpoint}}
\put(408,601){\usebox{\plotpoint}}
\put(409,602){\usebox{\plotpoint}}
\put(411,603){\usebox{\plotpoint}}
\put(412,604){\usebox{\plotpoint}}
\put(414,605){\usebox{\plotpoint}}
\put(416,606){\usebox{\plotpoint}}
\put(417,607){\usebox{\plotpoint}}
\put(419,608){\usebox{\plotpoint}}
\put(421,609){\usebox{\plotpoint}}
\put(422,610){\usebox{\plotpoint}}
\put(424,611){\usebox{\plotpoint}}
\put(426,612){\usebox{\plotpoint}}
\put(427,613){\usebox{\plotpoint}}
\put(429,614){\usebox{\plotpoint}}
\put(430,615){\usebox{\plotpoint}}
\put(432,616){\usebox{\plotpoint}}
\put(434,617){\usebox{\plotpoint}}
\put(436,618){\usebox{\plotpoint}}
\put(437,619){\usebox{\plotpoint}}
\put(439,620){\usebox{\plotpoint}}
\put(441,621){\usebox{\plotpoint}}
\put(443,622){\usebox{\plotpoint}}
\put(444,623){\usebox{\plotpoint}}
\put(446,624){\usebox{\plotpoint}}
\put(448,625){\usebox{\plotpoint}}
\put(449,626){\usebox{\plotpoint}}
\put(451,627){\usebox{\plotpoint}}
\put(453,628){\usebox{\plotpoint}}
\put(455,629){\usebox{\plotpoint}}
\put(457,630){\usebox{\plotpoint}}
\put(459,631){\usebox{\plotpoint}}
\put(461,632){\usebox{\plotpoint}}
\put(463,633){\usebox{\plotpoint}}
\put(465,634){\usebox{\plotpoint}}
\put(467,635){\usebox{\plotpoint}}
\put(469,636){\usebox{\plotpoint}}
\put(471,637){\usebox{\plotpoint}}
\put(473,638){\usebox{\plotpoint}}
\put(475,639){\usebox{\plotpoint}}
\put(477,640){\usebox{\plotpoint}}
\put(479,641){\usebox{\plotpoint}}
\put(481,642){\usebox{\plotpoint}}
\put(483,643){\usebox{\plotpoint}}
\put(485,644){\usebox{\plotpoint}}
\put(487,645){\usebox{\plotpoint}}
\put(489,646){\usebox{\plotpoint}}
\put(492,647){\usebox{\plotpoint}}
\put(494,648){\usebox{\plotpoint}}
\put(496,649){\usebox{\plotpoint}}
\put(498,650){\usebox{\plotpoint}}
\put(500,651){\usebox{\plotpoint}}
\put(503,652){\usebox{\plotpoint}}
\put(505,653){\usebox{\plotpoint}}
\put(507,654){\usebox{\plotpoint}}
\put(510,655){\usebox{\plotpoint}}
\put(512,656){\usebox{\plotpoint}}
\put(515,657){\usebox{\plotpoint}}
\put(517,658){\usebox{\plotpoint}}
\put(519,659){\usebox{\plotpoint}}
\put(522,660){\usebox{\plotpoint}}
\put(524,661){\usebox{\plotpoint}}
\put(527,662){\usebox{\plotpoint}}
\put(529,663){\usebox{\plotpoint}}
\put(532,664){\usebox{\plotpoint}}
\put(535,665){\usebox{\plotpoint}}
\put(537,666){\usebox{\plotpoint}}
\put(540,667){\usebox{\plotpoint}}
\put(543,668){\usebox{\plotpoint}}
\put(545,669){\usebox{\plotpoint}}
\put(548,670){\usebox{\plotpoint}}
\put(551,671){\usebox{\plotpoint}}
\put(554,672){\usebox{\plotpoint}}
\put(557,673){\usebox{\plotpoint}}
\put(560,674){\usebox{\plotpoint}}
\put(563,675){\usebox{\plotpoint}}
\put(566,676){\usebox{\plotpoint}}
\put(569,677){\usebox{\plotpoint}}
\put(572,678){\usebox{\plotpoint}}
\put(575,679){\usebox{\plotpoint}}
\put(578,680){\usebox{\plotpoint}}
\put(581,681){\usebox{\plotpoint}}
\put(585,682){\usebox{\plotpoint}}
\put(588,683){\usebox{\plotpoint}}
\put(592,684){\usebox{\plotpoint}}
\put(595,685){\usebox{\plotpoint}}
\put(599,686){\rule[-0.500pt]{1.445pt}{1.000pt}}
\put(605,687){\rule[-0.500pt]{1.445pt}{1.000pt}}
\put(611,688){\rule[-0.500pt]{1.445pt}{1.000pt}}
\put(617,689){\rule[-0.500pt]{1.445pt}{1.000pt}}
\put(623,690){\rule[-0.500pt]{1.927pt}{1.000pt}}
\put(631,691){\rule[-0.500pt]{1.927pt}{1.000pt}}
\put(639,692){\rule[-0.500pt]{1.927pt}{1.000pt}}
\put(647,693){\rule[-0.500pt]{5.782pt}{1.000pt}}
\put(671,694){\rule[-0.500pt]{1.927pt}{1.000pt}}
\put(679,693){\rule[-0.500pt]{1.927pt}{1.000pt}}
\put(687,692){\rule[-0.500pt]{1.927pt}{1.000pt}}
\put(695,691){\usebox{\plotpoint}}
\put(698,690){\usebox{\plotpoint}}
\put(702,689){\usebox{\plotpoint}}
\put(706,688){\usebox{\plotpoint}}
\put(710,687){\usebox{\plotpoint}}
\put(714,686){\usebox{\plotpoint}}
\put(717,685){\usebox{\plotpoint}}
\put(719,684){\usebox{\plotpoint}}
\put(721,683){\usebox{\plotpoint}}
\put(723,682){\usebox{\plotpoint}}
\put(725,681){\usebox{\plotpoint}}
\put(727,680){\usebox{\plotpoint}}
\put(729,679){\usebox{\plotpoint}}
\put(730,678){\usebox{\plotpoint}}
\put(732,677){\usebox{\plotpoint}}
\put(734,676){\usebox{\plotpoint}}
\put(736,675){\usebox{\plotpoint}}
\put(738,674){\usebox{\plotpoint}}
\put(740,673){\usebox{\plotpoint}}
\put(741,672){\usebox{\plotpoint}}
\put(743,671){\usebox{\plotpoint}}
\put(744,670){\usebox{\plotpoint}}
\put(745,669){\usebox{\plotpoint}}
\put(746,668){\usebox{\plotpoint}}
\put(748,667){\usebox{\plotpoint}}
\put(749,666){\usebox{\plotpoint}}
\put(750,665){\usebox{\plotpoint}}
\put(751,664){\usebox{\plotpoint}}
\put(752,663){\usebox{\plotpoint}}
\put(754,662){\usebox{\plotpoint}}
\put(755,661){\usebox{\plotpoint}}
\put(756,660){\usebox{\plotpoint}}
\put(757,659){\usebox{\plotpoint}}
\put(758,658){\usebox{\plotpoint}}
\put(760,657){\usebox{\plotpoint}}
\put(761,656){\usebox{\plotpoint}}
\put(762,655){\usebox{\plotpoint}}
\put(763,654){\usebox{\plotpoint}}
\put(764,653){\usebox{\plotpoint}}
\put(766,650){\usebox{\plotpoint}}
\put(767,649){\usebox{\plotpoint}}
\put(768,648){\usebox{\plotpoint}}
\put(769,646){\usebox{\plotpoint}}
\put(770,645){\usebox{\plotpoint}}
\put(771,644){\usebox{\plotpoint}}
\put(772,642){\usebox{\plotpoint}}
\put(773,641){\usebox{\plotpoint}}
\put(774,640){\usebox{\plotpoint}}
\put(775,638){\usebox{\plotpoint}}
\put(776,637){\usebox{\plotpoint}}
\put(777,636){\usebox{\plotpoint}}
\put(778,634){\usebox{\plotpoint}}
\put(779,633){\usebox{\plotpoint}}
\put(780,632){\usebox{\plotpoint}}
\put(781,630){\usebox{\plotpoint}}
\put(782,629){\usebox{\plotpoint}}
\put(783,628){\usebox{\plotpoint}}
\put(784,626){\usebox{\plotpoint}}
\put(785,625){\usebox{\plotpoint}}
\put(786,624){\usebox{\plotpoint}}
\put(787,622){\usebox{\plotpoint}}
\put(788,621){\usebox{\plotpoint}}
\put(789,620){\usebox{\plotpoint}}
\put(790,617){\usebox{\plotpoint}}
\put(791,615){\usebox{\plotpoint}}
\put(792,613){\usebox{\plotpoint}}
\put(793,611){\usebox{\plotpoint}}
\put(794,609){\usebox{\plotpoint}}
\put(795,607){\usebox{\plotpoint}}
\put(796,605){\usebox{\plotpoint}}
\put(797,603){\usebox{\plotpoint}}
\put(798,601){\usebox{\plotpoint}}
\put(799,599){\usebox{\plotpoint}}
\put(800,597){\usebox{\plotpoint}}
\put(801,595){\usebox{\plotpoint}}
\put(802,593){\usebox{\plotpoint}}
\put(803,591){\usebox{\plotpoint}}
\put(804,589){\usebox{\plotpoint}}
\put(805,587){\usebox{\plotpoint}}
\put(806,585){\usebox{\plotpoint}}
\put(807,583){\usebox{\plotpoint}}
\put(808,581){\usebox{\plotpoint}}
\put(809,579){\usebox{\plotpoint}}
\put(810,577){\usebox{\plotpoint}}
\put(811,575){\usebox{\plotpoint}}
\put(812,573){\usebox{\plotpoint}}
\put(813,571){\usebox{\plotpoint}}
\put(814,567){\usebox{\plotpoint}}
\put(815,564){\usebox{\plotpoint}}
\put(816,561){\usebox{\plotpoint}}
\put(817,558){\usebox{\plotpoint}}
\put(818,554){\usebox{\plotpoint}}
\put(819,551){\usebox{\plotpoint}}
\put(820,548){\usebox{\plotpoint}}
\put(821,545){\usebox{\plotpoint}}
\put(822,541){\usebox{\plotpoint}}
\put(823,538){\usebox{\plotpoint}}
\put(824,535){\usebox{\plotpoint}}
\put(825,532){\usebox{\plotpoint}}
\put(826,528){\usebox{\plotpoint}}
\put(827,525){\usebox{\plotpoint}}
\put(828,522){\usebox{\plotpoint}}
\put(829,519){\usebox{\plotpoint}}
\put(830,515){\usebox{\plotpoint}}
\put(831,512){\usebox{\plotpoint}}
\put(832,509){\usebox{\plotpoint}}
\put(833,506){\usebox{\plotpoint}}
\put(834,502){\usebox{\plotpoint}}
\put(835,499){\usebox{\plotpoint}}
\put(836,496){\usebox{\plotpoint}}
\put(837,493){\usebox{\plotpoint}}
\put(838,487){\rule[-0.500pt]{1.000pt}{1.265pt}}
\put(839,482){\rule[-0.500pt]{1.000pt}{1.265pt}}
\put(840,477){\rule[-0.500pt]{1.000pt}{1.265pt}}
\put(841,472){\rule[-0.500pt]{1.000pt}{1.265pt}}
\put(842,466){\rule[-0.500pt]{1.000pt}{1.265pt}}
\put(843,461){\rule[-0.500pt]{1.000pt}{1.265pt}}
\put(844,456){\rule[-0.500pt]{1.000pt}{1.265pt}}
\put(845,451){\rule[-0.500pt]{1.000pt}{1.265pt}}
\put(846,445){\rule[-0.500pt]{1.000pt}{1.265pt}}
\put(847,440){\rule[-0.500pt]{1.000pt}{1.265pt}}
\put(848,435){\rule[-0.500pt]{1.000pt}{1.265pt}}
\put(849,430){\rule[-0.500pt]{1.000pt}{1.265pt}}
\put(850,424){\rule[-0.500pt]{1.000pt}{1.265pt}}
\put(851,419){\rule[-0.500pt]{1.000pt}{1.265pt}}
\put(852,414){\rule[-0.500pt]{1.000pt}{1.265pt}}
\put(853,409){\rule[-0.500pt]{1.000pt}{1.265pt}}
\put(854,403){\rule[-0.500pt]{1.000pt}{1.265pt}}
\put(855,398){\rule[-0.500pt]{1.000pt}{1.265pt}}
\put(856,393){\rule[-0.500pt]{1.000pt}{1.265pt}}
\put(857,388){\rule[-0.500pt]{1.000pt}{1.265pt}}
\put(858,382){\rule[-0.500pt]{1.000pt}{1.265pt}}
\put(859,377){\rule[-0.500pt]{1.000pt}{1.265pt}}
\put(860,372){\rule[-0.500pt]{1.000pt}{1.265pt}}
\put(861,367){\rule[-0.500pt]{1.000pt}{1.265pt}}
\put(862,357){\rule[-0.500pt]{1.000pt}{2.228pt}}
\put(863,348){\rule[-0.500pt]{1.000pt}{2.228pt}}
\put(864,339){\rule[-0.500pt]{1.000pt}{2.228pt}}
\put(865,330){\rule[-0.500pt]{1.000pt}{2.228pt}}
\put(866,320){\rule[-0.500pt]{1.000pt}{2.228pt}}
\put(867,311){\rule[-0.500pt]{1.000pt}{2.228pt}}
\put(868,302){\rule[-0.500pt]{1.000pt}{2.228pt}}
\put(869,293){\rule[-0.500pt]{1.000pt}{2.228pt}}
\put(870,283){\rule[-0.500pt]{1.000pt}{2.228pt}}
\put(871,274){\rule[-0.500pt]{1.000pt}{2.228pt}}
\put(872,265){\rule[-0.500pt]{1.000pt}{2.228pt}}
\put(873,256){\rule[-0.500pt]{1.000pt}{2.228pt}}
\put(874,246){\rule[-0.500pt]{1.000pt}{2.228pt}}
\put(875,237){\rule[-0.500pt]{1.000pt}{2.228pt}}
\put(876,228){\rule[-0.500pt]{1.000pt}{2.228pt}}
\put(877,219){\rule[-0.500pt]{1.000pt}{2.228pt}}
\put(878,209){\rule[-0.500pt]{1.000pt}{2.228pt}}
\put(879,200){\rule[-0.500pt]{1.000pt}{2.228pt}}
\put(880,191){\rule[-0.500pt]{1.000pt}{2.228pt}}
\put(881,182){\rule[-0.500pt]{1.000pt}{2.228pt}}
\put(882,172){\rule[-0.500pt]{1.000pt}{2.228pt}}
\put(883,163){\rule[-0.500pt]{1.000pt}{2.228pt}}
\put(884,154){\rule[-0.500pt]{1.000pt}{2.228pt}}
\put(885,145){\rule[-0.500pt]{1.000pt}{2.228pt}}
\sbox{\plotpoint}{\rule[-1.000pt]{2.000pt}{2.000pt}}%
\put(264,370){\usebox{\plotpoint}}
\put(264,370){\usebox{\plotpoint}}
\put(265,373){\usebox{\plotpoint}}
\put(266,376){\usebox{\plotpoint}}
\put(267,379){\usebox{\plotpoint}}
\put(268,382){\usebox{\plotpoint}}
\put(269,386){\usebox{\plotpoint}}
\put(270,389){\usebox{\plotpoint}}
\put(271,392){\usebox{\plotpoint}}
\put(272,395){\usebox{\plotpoint}}
\put(273,398){\usebox{\plotpoint}}
\put(274,402){\usebox{\plotpoint}}
\put(275,405){\usebox{\plotpoint}}
\put(276,408){\usebox{\plotpoint}}
\put(277,411){\usebox{\plotpoint}}
\put(278,414){\usebox{\plotpoint}}
\put(279,418){\usebox{\plotpoint}}
\put(280,421){\usebox{\plotpoint}}
\put(281,424){\usebox{\plotpoint}}
\put(282,427){\usebox{\plotpoint}}
\put(283,430){\usebox{\plotpoint}}
\put(284,434){\usebox{\plotpoint}}
\put(285,437){\usebox{\plotpoint}}
\put(286,440){\usebox{\plotpoint}}
\put(287,443){\usebox{\plotpoint}}
\put(288,447){\usebox{\plotpoint}}
\put(289,448){\usebox{\plotpoint}}
\put(290,449){\usebox{\plotpoint}}
\put(291,450){\usebox{\plotpoint}}
\put(292,452){\usebox{\plotpoint}}
\put(293,453){\usebox{\plotpoint}}
\put(294,454){\usebox{\plotpoint}}
\put(295,455){\usebox{\plotpoint}}
\put(296,457){\usebox{\plotpoint}}
\put(297,458){\usebox{\plotpoint}}
\put(298,459){\usebox{\plotpoint}}
\put(299,460){\usebox{\plotpoint}}
\put(300,462){\usebox{\plotpoint}}
\put(301,463){\usebox{\plotpoint}}
\put(302,464){\usebox{\plotpoint}}
\put(303,465){\usebox{\plotpoint}}
\put(304,467){\usebox{\plotpoint}}
\put(305,468){\usebox{\plotpoint}}
\put(306,469){\usebox{\plotpoint}}
\put(307,470){\usebox{\plotpoint}}
\put(308,472){\usebox{\plotpoint}}
\put(309,473){\usebox{\plotpoint}}
\put(310,474){\usebox{\plotpoint}}
\put(311,475){\usebox{\plotpoint}}
\put(312,477){\usebox{\plotpoint}}
\put(313,478){\usebox{\plotpoint}}
\put(314,479){\usebox{\plotpoint}}
\put(315,480){\usebox{\plotpoint}}
\put(316,481){\usebox{\plotpoint}}
\put(317,482){\usebox{\plotpoint}}
\put(318,483){\usebox{\plotpoint}}
\put(319,484){\usebox{\plotpoint}}
\put(321,485){\usebox{\plotpoint}}
\put(322,486){\usebox{\plotpoint}}
\put(323,487){\usebox{\plotpoint}}
\put(324,488){\usebox{\plotpoint}}
\put(325,489){\usebox{\plotpoint}}
\put(326,490){\usebox{\plotpoint}}
\put(327,491){\usebox{\plotpoint}}
\put(329,492){\usebox{\plotpoint}}
\put(330,493){\usebox{\plotpoint}}
\put(331,494){\usebox{\plotpoint}}
\put(332,495){\usebox{\plotpoint}}
\put(333,496){\usebox{\plotpoint}}
\put(334,497){\usebox{\plotpoint}}
\put(335,498){\usebox{\plotpoint}}
\put(337,499){\usebox{\plotpoint}}
\put(338,500){\usebox{\plotpoint}}
\put(340,501){\usebox{\plotpoint}}
\put(341,502){\usebox{\plotpoint}}
\put(343,503){\usebox{\plotpoint}}
\put(344,504){\usebox{\plotpoint}}
\put(345,505){\usebox{\plotpoint}}
\put(347,506){\usebox{\plotpoint}}
\put(348,507){\usebox{\plotpoint}}
\put(350,508){\usebox{\plotpoint}}
\put(351,509){\usebox{\plotpoint}}
\put(352,510){\usebox{\plotpoint}}
\put(354,511){\usebox{\plotpoint}}
\put(355,512){\usebox{\plotpoint}}
\put(357,513){\usebox{\plotpoint}}
\put(358,514){\usebox{\plotpoint}}
\put(360,515){\usebox{\plotpoint}}
\put(361,516){\usebox{\plotpoint}}
\put(363,517){\usebox{\plotpoint}}
\put(364,518){\usebox{\plotpoint}}
\put(366,519){\usebox{\plotpoint}}
\put(368,520){\usebox{\plotpoint}}
\put(369,521){\usebox{\plotpoint}}
\put(371,522){\usebox{\plotpoint}}
\put(372,523){\usebox{\plotpoint}}
\put(374,524){\usebox{\plotpoint}}
\put(376,525){\usebox{\plotpoint}}
\put(377,526){\usebox{\plotpoint}}
\put(379,527){\usebox{\plotpoint}}
\put(380,528){\usebox{\plotpoint}}
\put(382,529){\usebox{\plotpoint}}
\put(384,530){\usebox{\plotpoint}}
\put(385,531){\usebox{\plotpoint}}
\put(387,532){\usebox{\plotpoint}}
\put(389,533){\usebox{\plotpoint}}
\put(391,534){\usebox{\plotpoint}}
\put(393,535){\usebox{\plotpoint}}
\put(395,536){\usebox{\plotpoint}}
\put(396,537){\usebox{\plotpoint}}
\put(398,538){\usebox{\plotpoint}}
\put(400,539){\usebox{\plotpoint}}
\put(402,540){\usebox{\plotpoint}}
\put(404,541){\usebox{\plotpoint}}
\put(406,542){\usebox{\plotpoint}}
\put(408,543){\usebox{\plotpoint}}
\put(409,544){\usebox{\plotpoint}}
\put(411,545){\usebox{\plotpoint}}
\put(413,546){\usebox{\plotpoint}}
\put(415,547){\usebox{\plotpoint}}
\put(417,548){\usebox{\plotpoint}}
\put(419,549){\usebox{\plotpoint}}
\put(421,550){\usebox{\plotpoint}}
\put(423,551){\usebox{\plotpoint}}
\put(425,552){\usebox{\plotpoint}}
\put(427,553){\usebox{\plotpoint}}
\put(429,554){\usebox{\plotpoint}}
\put(430,555){\usebox{\plotpoint}}
\put(433,556){\usebox{\plotpoint}}
\put(435,557){\usebox{\plotpoint}}
\put(438,558){\usebox{\plotpoint}}
\put(440,559){\usebox{\plotpoint}}
\put(442,560){\usebox{\plotpoint}}
\put(445,561){\usebox{\plotpoint}}
\put(447,562){\usebox{\plotpoint}}
\put(450,563){\usebox{\plotpoint}}
\put(452,564){\usebox{\plotpoint}}
\put(454,565){\usebox{\plotpoint}}
\put(457,566){\usebox{\plotpoint}}
\put(459,567){\usebox{\plotpoint}}
\put(461,568){\usebox{\plotpoint}}
\put(463,569){\usebox{\plotpoint}}
\put(465,570){\usebox{\plotpoint}}
\put(468,571){\usebox{\plotpoint}}
\put(470,572){\usebox{\plotpoint}}
\put(472,573){\usebox{\plotpoint}}
\put(474,574){\usebox{\plotpoint}}
\put(476,575){\usebox{\plotpoint}}
\put(479,576){\usebox{\plotpoint}}
\put(481,577){\usebox{\plotpoint}}
\put(484,578){\usebox{\plotpoint}}
\put(486,579){\usebox{\plotpoint}}
\put(489,580){\usebox{\plotpoint}}
\put(492,581){\usebox{\plotpoint}}
\put(494,582){\usebox{\plotpoint}}
\put(497,583){\usebox{\plotpoint}}
\put(500,584){\usebox{\plotpoint}}
\put(502,585){\usebox{\plotpoint}}
\put(505,586){\usebox{\plotpoint}}
\put(508,587){\usebox{\plotpoint}}
\put(510,588){\usebox{\plotpoint}}
\put(513,589){\usebox{\plotpoint}}
\put(516,590){\usebox{\plotpoint}}
\put(519,591){\usebox{\plotpoint}}
\put(521,592){\usebox{\plotpoint}}
\put(524,593){\usebox{\plotpoint}}
\put(527,594){\usebox{\plotpoint}}
\put(529,595){\usebox{\plotpoint}}
\put(532,596){\usebox{\plotpoint}}
\put(535,597){\usebox{\plotpoint}}
\put(537,598){\usebox{\plotpoint}}
\put(540,599){\usebox{\plotpoint}}
\put(543,600){\usebox{\plotpoint}}
\put(545,601){\usebox{\plotpoint}}
\put(548,602){\usebox{\plotpoint}}
\put(551,603){\usebox{\plotpoint}}
\put(554,604){\usebox{\plotpoint}}
\put(557,605){\usebox{\plotpoint}}
\put(560,606){\usebox{\plotpoint}}
\put(563,607){\usebox{\plotpoint}}
\put(566,608){\usebox{\plotpoint}}
\put(569,609){\usebox{\plotpoint}}
\put(572,610){\usebox{\plotpoint}}
\put(575,611){\usebox{\plotpoint}}
\put(578,612){\usebox{\plotpoint}}
\put(581,613){\usebox{\plotpoint}}
\put(585,614){\usebox{\plotpoint}}
\put(588,615){\usebox{\plotpoint}}
\put(592,616){\usebox{\plotpoint}}
\put(595,617){\usebox{\plotpoint}}
\put(599,618){\usebox{\plotpoint}}
\put(602,619){\usebox{\plotpoint}}
\put(605,620){\usebox{\plotpoint}}
\put(608,621){\usebox{\plotpoint}}
\put(611,622){\usebox{\plotpoint}}
\put(614,623){\usebox{\plotpoint}}
\put(617,624){\usebox{\plotpoint}}
\put(620,625){\usebox{\plotpoint}}
\put(623,626){\usebox{\plotpoint}}
\put(627,627){\usebox{\plotpoint}}
\put(631,628){\usebox{\plotpoint}}
\put(635,629){\usebox{\plotpoint}}
\put(639,630){\usebox{\plotpoint}}
\put(643,631){\usebox{\plotpoint}}
\put(647,632){\usebox{\plotpoint}}
\put(650,633){\usebox{\plotpoint}}
\put(653,634){\usebox{\plotpoint}}
\put(657,635){\usebox{\plotpoint}}
\put(660,636){\usebox{\plotpoint}}
\put(664,637){\usebox{\plotpoint}}
\put(667,638){\usebox{\plotpoint}}
\put(671,639){\usebox{\plotpoint}}
\put(675,640){\usebox{\plotpoint}}
\put(679,641){\usebox{\plotpoint}}
\put(683,642){\usebox{\plotpoint}}
\put(687,643){\usebox{\plotpoint}}
\put(691,644){\usebox{\plotpoint}}
\put(695,645){\usebox{\plotpoint}}
\put(698,646){\usebox{\plotpoint}}
\put(702,647){\usebox{\plotpoint}}
\put(706,648){\usebox{\plotpoint}}
\put(710,649){\usebox{\plotpoint}}
\put(714,650){\usebox{\plotpoint}}
\put(717,651){\usebox{\plotpoint}}
\put(722,652){\usebox{\plotpoint}}
\put(726,653){\usebox{\plotpoint}}
\put(730,654){\usebox{\plotpoint}}
\put(734,655){\usebox{\plotpoint}}
\put(738,656){\usebox{\plotpoint}}
\put(742,657){\usebox{\plotpoint}}
\put(746,658){\usebox{\plotpoint}}
\put(750,659){\usebox{\plotpoint}}
\put(754,660){\usebox{\plotpoint}}
\put(758,661){\usebox{\plotpoint}}
\put(762,662){\usebox{\plotpoint}}
\put(766,663){\usebox{\plotpoint}}
\put(770,664){\usebox{\plotpoint}}
\put(775,665){\usebox{\plotpoint}}
\put(780,666){\usebox{\plotpoint}}
\put(785,667){\usebox{\plotpoint}}
\put(789,668){\usebox{\plotpoint}}
\put(796,669){\usebox{\plotpoint}}
\put(802,670){\usebox{\plotpoint}}
\put(808,671){\usebox{\plotpoint}}
\put(814,672){\usebox{\plotpoint}}
\put(818,673){\usebox{\plotpoint}}
\put(823,674){\usebox{\plotpoint}}
\put(828,675){\usebox{\plotpoint}}
\put(833,676){\usebox{\plotpoint}}
\put(837,677){\usebox{\plotpoint}}
\put(844,678){\usebox{\plotpoint}}
\put(850,679){\usebox{\plotpoint}}
\put(856,680){\usebox{\plotpoint}}
\put(862,681){\usebox{\plotpoint}}
\put(870,682){\usebox{\plotpoint}}
\put(878,683){\usebox{\plotpoint}}
\put(886,684){\usebox{\plotpoint}}
\put(894,685){\usebox{\plotpoint}}
\put(902,686){\usebox{\plotpoint}}
\put(910,687){\usebox{\plotpoint}}
\put(918,688){\usebox{\plotpoint}}
\put(926,689){\usebox{\plotpoint}}
\put(934,690){\rule[-1.000pt]{2.891pt}{2.000pt}}
\put(946,691){\rule[-1.000pt]{2.891pt}{2.000pt}}
\put(958,692){\rule[-1.000pt]{5.782pt}{2.000pt}}
\put(982,693){\rule[-1.000pt]{5.541pt}{2.000pt}}
\put(1005,694){\rule[-1.000pt]{11.563pt}{2.000pt}}
\put(1053,693){\rule[-1.000pt]{2.891pt}{2.000pt}}
\put(1065,692){\rule[-1.000pt]{2.891pt}{2.000pt}}
\put(1077,691){\usebox{\plotpoint}}
\put(1083,690){\usebox{\plotpoint}}
\put(1089,689){\usebox{\plotpoint}}
\put(1095,688){\usebox{\plotpoint}}
\put(1101,687){\usebox{\plotpoint}}
\put(1105,686){\usebox{\plotpoint}}
\put(1110,685){\usebox{\plotpoint}}
\put(1115,684){\usebox{\plotpoint}}
\put(1120,683){\usebox{\plotpoint}}
\put(1125,682){\usebox{\plotpoint}}
\put(1128,681){\usebox{\plotpoint}}
\put(1131,680){\usebox{\plotpoint}}
\put(1135,679){\usebox{\plotpoint}}
\put(1138,678){\usebox{\plotpoint}}
\put(1142,677){\usebox{\plotpoint}}
\put(1145,676){\usebox{\plotpoint}}
\put(1149,675){\usebox{\plotpoint}}
\put(1151,674){\usebox{\plotpoint}}
\put(1154,673){\usebox{\plotpoint}}
\put(1156,672){\usebox{\plotpoint}}
\put(1159,671){\usebox{\plotpoint}}
\put(1162,670){\usebox{\plotpoint}}
\put(1164,669){\usebox{\plotpoint}}
\put(1167,668){\usebox{\plotpoint}}
\put(1170,667){\usebox{\plotpoint}}
\put(1172,666){\usebox{\plotpoint}}
\put(1175,665){\usebox{\plotpoint}}
\put(1177,664){\usebox{\plotpoint}}
\put(1179,663){\usebox{\plotpoint}}
\put(1181,662){\usebox{\plotpoint}}
\put(1183,661){\usebox{\plotpoint}}
\put(1185,660){\usebox{\plotpoint}}
\put(1187,659){\usebox{\plotpoint}}
\put(1189,658){\usebox{\plotpoint}}
\put(1191,657){\usebox{\plotpoint}}
\put(1193,656){\usebox{\plotpoint}}
\put(1195,655){\usebox{\plotpoint}}
\put(1197,654){\usebox{\plotpoint}}
\put(1198,653){\usebox{\plotpoint}}
\put(1200,652){\usebox{\plotpoint}}
\put(1201,651){\usebox{\plotpoint}}
\put(1203,650){\usebox{\plotpoint}}
\put(1204,649){\usebox{\plotpoint}}
\put(1206,648){\usebox{\plotpoint}}
\put(1208,647){\usebox{\plotpoint}}
\put(1209,646){\usebox{\plotpoint}}
\put(1211,645){\usebox{\plotpoint}}
\put(1212,644){\usebox{\plotpoint}}
\put(1214,643){\usebox{\plotpoint}}
\put(1216,642){\usebox{\plotpoint}}
\put(1217,641){\usebox{\plotpoint}}
\put(1219,640){\usebox{\plotpoint}}
\put(1220,639){\usebox{\plotpoint}}
\put(1222,638){\usebox{\plotpoint}}
\put(1223,637){\usebox{\plotpoint}}
\put(1224,636){\usebox{\plotpoint}}
\put(1226,635){\usebox{\plotpoint}}
\put(1227,634){\usebox{\plotpoint}}
\put(1228,633){\usebox{\plotpoint}}
\put(1229,632){\usebox{\plotpoint}}
\put(1231,631){\usebox{\plotpoint}}
\put(1232,630){\usebox{\plotpoint}}
\put(1233,629){\usebox{\plotpoint}}
\put(1234,628){\usebox{\plotpoint}}
\put(1236,627){\usebox{\plotpoint}}
\put(1237,626){\usebox{\plotpoint}}
\put(1238,625){\usebox{\plotpoint}}
\put(1239,624){\usebox{\plotpoint}}
\put(1241,623){\usebox{\plotpoint}}
\put(1242,622){\usebox{\plotpoint}}
\put(1243,621){\usebox{\plotpoint}}
\put(1245,618){\usebox{\plotpoint}}
\put(1246,617){\usebox{\plotpoint}}
\put(1247,616){\usebox{\plotpoint}}
\put(1248,615){\usebox{\plotpoint}}
\put(1249,614){\usebox{\plotpoint}}
\put(1250,613){\usebox{\plotpoint}}
\put(1251,612){\usebox{\plotpoint}}
\put(1252,611){\usebox{\plotpoint}}
\put(1253,610){\usebox{\plotpoint}}
\put(1254,609){\usebox{\plotpoint}}
\put(1255,608){\usebox{\plotpoint}}
\put(1256,607){\usebox{\plotpoint}}
\put(1257,606){\usebox{\plotpoint}}
\put(1258,605){\usebox{\plotpoint}}
\put(1259,604){\usebox{\plotpoint}}
\put(1260,603){\usebox{\plotpoint}}
\put(1261,602){\usebox{\plotpoint}}
\put(1262,601){\usebox{\plotpoint}}
\put(1263,600){\usebox{\plotpoint}}
\put(1264,599){\usebox{\plotpoint}}
\put(1265,598){\usebox{\plotpoint}}
\put(1266,597){\usebox{\plotpoint}}
\put(1267,596){\usebox{\plotpoint}}
\put(1268,595){\usebox{\plotpoint}}
\put(1269,593){\usebox{\plotpoint}}
\put(1270,592){\usebox{\plotpoint}}
\put(1271,590){\usebox{\plotpoint}}
\put(1272,589){\usebox{\plotpoint}}
\put(1273,588){\usebox{\plotpoint}}
\put(1274,586){\usebox{\plotpoint}}
\put(1275,585){\usebox{\plotpoint}}
\put(1276,584){\usebox{\plotpoint}}
\put(1277,582){\usebox{\plotpoint}}
\put(1278,581){\usebox{\plotpoint}}
\put(1279,580){\usebox{\plotpoint}}
\put(1280,578){\usebox{\plotpoint}}
\put(1281,577){\usebox{\plotpoint}}
\put(1282,576){\usebox{\plotpoint}}
\put(1283,574){\usebox{\plotpoint}}
\put(1284,573){\usebox{\plotpoint}}
\put(1285,572){\usebox{\plotpoint}}
\put(1286,570){\usebox{\plotpoint}}
\put(1287,569){\usebox{\plotpoint}}
\put(1288,568){\usebox{\plotpoint}}
\put(1289,566){\usebox{\plotpoint}}
\put(1290,565){\usebox{\plotpoint}}
\put(1291,564){\usebox{\plotpoint}}
\put(1292,562){\usebox{\plotpoint}}
\put(1293,560){\usebox{\plotpoint}}
\put(1294,559){\usebox{\plotpoint}}
\put(1295,557){\usebox{\plotpoint}}
\put(1296,555){\usebox{\plotpoint}}
\put(1297,554){\usebox{\plotpoint}}
\put(1298,552){\usebox{\plotpoint}}
\put(1299,551){\usebox{\plotpoint}}
\put(1300,549){\usebox{\plotpoint}}
\put(1301,547){\usebox{\plotpoint}}
\put(1302,546){\usebox{\plotpoint}}
\put(1303,544){\usebox{\plotpoint}}
\put(1304,542){\usebox{\plotpoint}}
\put(1305,541){\usebox{\plotpoint}}
\put(1306,539){\usebox{\plotpoint}}
\put(1307,538){\usebox{\plotpoint}}
\put(1308,536){\usebox{\plotpoint}}
\put(1309,534){\usebox{\plotpoint}}
\put(1310,533){\usebox{\plotpoint}}
\put(1311,531){\usebox{\plotpoint}}
\put(1312,529){\usebox{\plotpoint}}
\put(1313,528){\usebox{\plotpoint}}
\put(1314,526){\usebox{\plotpoint}}
\put(1315,525){\usebox{\plotpoint}}
\put(1316,522){\usebox{\plotpoint}}
\put(1317,520){\usebox{\plotpoint}}
\put(1318,518){\usebox{\plotpoint}}
\put(1319,516){\usebox{\plotpoint}}
\put(1320,514){\usebox{\plotpoint}}
\put(1321,511){\usebox{\plotpoint}}
\put(1322,509){\usebox{\plotpoint}}
\put(1323,507){\usebox{\plotpoint}}
\put(1324,505){\usebox{\plotpoint}}
\put(1325,503){\usebox{\plotpoint}}
\put(1326,501){\usebox{\plotpoint}}
\put(1327,498){\usebox{\plotpoint}}
\put(1328,496){\usebox{\plotpoint}}
\put(1329,494){\usebox{\plotpoint}}
\put(1330,492){\usebox{\plotpoint}}
\put(1331,490){\usebox{\plotpoint}}
\put(1332,488){\usebox{\plotpoint}}
\put(1333,486){\usebox{\plotpoint}}
\put(1334,483){\usebox{\plotpoint}}
\put(1335,481){\usebox{\plotpoint}}
\put(1336,479){\usebox{\plotpoint}}
\put(1337,477){\usebox{\plotpoint}}
\put(1338,475){\usebox{\plotpoint}}
\put(1339,473){\usebox{\plotpoint}}
\put(1340,470){\usebox{\plotpoint}}
\put(1341,467){\usebox{\plotpoint}}
\put(1342,464){\usebox{\plotpoint}}
\put(1343,461){\usebox{\plotpoint}}
\put(1344,459){\usebox{\plotpoint}}
\put(1345,456){\usebox{\plotpoint}}
\put(1346,453){\usebox{\plotpoint}}
\put(1347,450){\usebox{\plotpoint}}
\put(1348,447){\usebox{\plotpoint}}
\put(1349,445){\usebox{\plotpoint}}
\put(1350,442){\usebox{\plotpoint}}
\put(1351,439){\usebox{\plotpoint}}
\put(1352,436){\usebox{\plotpoint}}
\put(1353,433){\usebox{\plotpoint}}
\put(1354,431){\usebox{\plotpoint}}
\put(1355,428){\usebox{\plotpoint}}
\put(1356,425){\usebox{\plotpoint}}
\put(1357,422){\usebox{\plotpoint}}
\put(1358,419){\usebox{\plotpoint}}
\put(1359,417){\usebox{\plotpoint}}
\put(1360,414){\usebox{\plotpoint}}
\put(1361,411){\usebox{\plotpoint}}
\put(1362,408){\usebox{\plotpoint}}
\put(1363,406){\usebox{\plotpoint}}
\put(1364,402){\usebox{\plotpoint}}
\put(1365,398){\usebox{\plotpoint}}
\put(1366,394){\usebox{\plotpoint}}
\put(1367,390){\usebox{\plotpoint}}
\put(1368,387){\usebox{\plotpoint}}
\put(1369,383){\usebox{\plotpoint}}
\put(1370,379){\usebox{\plotpoint}}
\put(1371,375){\usebox{\plotpoint}}
\put(1372,371){\usebox{\plotpoint}}
\put(1373,368){\usebox{\plotpoint}}
\put(1374,364){\usebox{\plotpoint}}
\put(1375,360){\usebox{\plotpoint}}
\put(1376,356){\usebox{\plotpoint}}
\put(1377,352){\usebox{\plotpoint}}
\put(1378,349){\usebox{\plotpoint}}
\put(1379,345){\usebox{\plotpoint}}
\put(1380,341){\usebox{\plotpoint}}
\put(1381,337){\usebox{\plotpoint}}
\put(1382,333){\usebox{\plotpoint}}
\put(1383,330){\usebox{\plotpoint}}
\put(1384,326){\usebox{\plotpoint}}
\put(1385,322){\usebox{\plotpoint}}
\put(1386,318){\usebox{\plotpoint}}
\put(1387,315){\usebox{\plotpoint}}
\put(1388,309){\usebox{\plotpoint}}
\put(1389,304){\usebox{\plotpoint}}
\put(1390,299){\usebox{\plotpoint}}
\put(1391,294){\usebox{\plotpoint}}
\put(1392,289){\usebox{\plotpoint}}
\put(1393,284){\usebox{\plotpoint}}
\put(1394,278){\usebox{\plotpoint}}
\put(1395,273){\usebox{\plotpoint}}
\put(1396,268){\usebox{\plotpoint}}
\put(1397,263){\usebox{\plotpoint}}
\put(1398,258){\usebox{\plotpoint}}
\put(1399,253){\usebox{\plotpoint}}
\put(1400,247){\usebox{\plotpoint}}
\put(1401,242){\usebox{\plotpoint}}
\put(1402,237){\usebox{\plotpoint}}
\put(1403,232){\usebox{\plotpoint}}
\put(1404,227){\usebox{\plotpoint}}
\put(1405,222){\usebox{\plotpoint}}
\put(1406,216){\usebox{\plotpoint}}
\put(1407,211){\usebox{\plotpoint}}
\put(1408,206){\usebox{\plotpoint}}
\put(1409,201){\usebox{\plotpoint}}
\put(1410,196){\usebox{\plotpoint}}
\put(1411,191){\usebox{\plotpoint}}
\put(1412,183){\usebox{\plotpoint}}
\put(1413,176){\usebox{\plotpoint}}
\put(1414,169){\usebox{\plotpoint}}
\put(1415,162){\usebox{\plotpoint}}
\put(1416,155){\usebox{\plotpoint}}
\put(1417,148){\usebox{\plotpoint}}
\put(1418,141){\usebox{\plotpoint}}
\put(1419,134){\usebox{\plotpoint}}
\put(1420,127){\usebox{\plotpoint}}
\put(1421,120){\usebox{\plotpoint}}
\put(1422,113){\usebox{\plotpoint}}
\end{picture}

%% file: fig3b.tex
% GNUPLOT: LaTeX picture
\setlength{\unitlength}{0.240900pt}
\begin{picture}(1500,900)(0,0)
\tenrm
\ifx\plotpoint\undefined\newsavebox{\plotpoint}\fi
\put(264,113){\line(1,0){1172}}
\put(264,113){\line(1,0){20}}
\put(1436,113){\line(-1,0){20}}
\put(242,113){\makebox(0,0)[r]{0}}
\put(264,257){\line(1,0){20}}
\put(1436,257){\line(-1,0){20}}
\put(242,257){\makebox(0,0)[r]{0.2}}
\put(264,401){\line(1,0){20}}
\put(1436,401){\line(-1,0){20}}
\put(242,401){\makebox(0,0)[r]{0.4}}
\put(264,544){\line(1,0){20}}
\put(1436,544){\line(-1,0){20}}
\put(242,544){\makebox(0,0)[r]{0.6}}
\put(264,688){\line(1,0){20}}
\put(1436,688){\line(-1,0){20}}
\put(242,688){\makebox(0,0)[r]{0.8}}
\put(264,832){\line(1,0){20}}
\put(1436,832){\line(-1,0){20}}
\put(242,832){\makebox(0,0)[r]{1}}
\put(264,113){\line(0,1){20}}
\put(264,832){\line(0,-1){20}}
\put(264,68){\makebox(0,0){50}}
\put(382,113){\line(0,1){20}}
\put(382,832){\line(0,-1){20}}
\put(382,68){\makebox(0,0){550}}
\put(501,113){\line(0,1){20}}
\put(501,832){\line(0,-1){20}}
\put(501,68){\makebox(0,0){1050}}
\put(619,113){\line(0,1){20}}
\put(619,832){\line(0,-1){20}}
\put(619,68){\makebox(0,0){1550}}
\put(738,113){\line(0,1){20}}
\put(738,832){\line(0,-1){20}}
\put(738,68){\makebox(0,0){2050}}
\put(856,113){\line(0,1){20}}
\put(856,832){\line(0,-1){20}}
\put(856,68){\makebox(0,0){2550}}
\put(974,113){\line(0,1){20}}
\put(974,832){\line(0,-1){20}}
\put(974,68){\makebox(0,0){3050}}
\put(1093,113){\line(0,1){20}}
\put(1093,832){\line(0,-1){20}}
\put(1093,68){\makebox(0,0){3550}}
\put(1211,113){\line(0,1){20}}
\put(1211,832){\line(0,-1){20}}
\put(1211,68){\makebox(0,0){4050}}
\put(1329,113){\line(0,1){20}}
\put(1329,832){\line(0,-1){20}}
\put(1329,68){\makebox(0,0){4550}}
\put(264,113){\line(1,0){1172}}
\put(1436,113){\line(0,1){719}}
\put(1436,832){\line(-1,0){1172}}
\put(264,832){\line(0,-1){719}}
\put(45,472){\makebox(0,0)[l]{\shortstack{$A(m_{\tilde{w}})$}}}
\put(850,23){\makebox(0,0){$m_{\tilde{w}}$}}
\put(962,760){\makebox(0,0)[l]{$m_{\tilde{q}}=0.25 TeV$}}
\sbox{\plotpoint}{\rule[-0.200pt]{0.400pt}{0.400pt}}%
\put(264,306){\usebox{\plotpoint}}
\put(264,306){\usebox{\plotpoint}}
\put(265,307){\usebox{\plotpoint}}
\put(266,308){\usebox{\plotpoint}}
\put(267,310){\usebox{\plotpoint}}
\put(268,311){\usebox{\plotpoint}}
\put(269,312){\usebox{\plotpoint}}
\put(270,314){\usebox{\plotpoint}}
\put(271,315){\usebox{\plotpoint}}
\put(272,316){\usebox{\plotpoint}}
\put(273,317){\usebox{\plotpoint}}
\put(274,318){\usebox{\plotpoint}}
\put(275,319){\usebox{\plotpoint}}
\put(276,320){\usebox{\plotpoint}}
\put(277,321){\usebox{\plotpoint}}
\put(278,322){\usebox{\plotpoint}}
\put(279,323){\usebox{\plotpoint}}
\put(280,324){\usebox{\plotpoint}}
\put(282,325){\rule[-0.200pt]{0.482pt}{0.400pt}}
\put(284,326){\rule[-0.200pt]{0.482pt}{0.400pt}}
\put(286,327){\rule[-0.200pt]{0.482pt}{0.400pt}}
\put(288,328){\rule[-0.200pt]{0.482pt}{0.400pt}}
\put(290,329){\rule[-0.200pt]{0.482pt}{0.400pt}}
\put(292,330){\rule[-0.200pt]{0.482pt}{0.400pt}}
\put(294,331){\rule[-0.200pt]{0.723pt}{0.400pt}}
\put(297,332){\rule[-0.200pt]{0.723pt}{0.400pt}}
\put(300,333){\rule[-0.200pt]{0.401pt}{0.400pt}}
\put(301,334){\rule[-0.200pt]{0.401pt}{0.400pt}}
\put(303,335){\rule[-0.200pt]{0.401pt}{0.400pt}}
\put(304,336){\rule[-0.200pt]{0.482pt}{0.400pt}}
\put(307,337){\rule[-0.200pt]{0.482pt}{0.400pt}}
\put(309,338){\rule[-0.200pt]{0.482pt}{0.400pt}}
\put(311,339){\usebox{\plotpoint}}
\put(312,340){\usebox{\plotpoint}}
\put(314,341){\usebox{\plotpoint}}
\put(315,342){\usebox{\plotpoint}}
\put(317,343){\usebox{\plotpoint}}
\put(318,344){\usebox{\plotpoint}}
\put(320,345){\usebox{\plotpoint}}
\put(321,346){\usebox{\plotpoint}}
\put(323,347){\usebox{\plotpoint}}
\put(324,348){\usebox{\plotpoint}}
\put(325,349){\usebox{\plotpoint}}
\put(326,350){\usebox{\plotpoint}}
\put(327,351){\usebox{\plotpoint}}
\put(329,352){\usebox{\plotpoint}}
\put(330,353){\usebox{\plotpoint}}
\put(331,354){\usebox{\plotpoint}}
\put(332,355){\usebox{\plotpoint}}
\put(333,356){\usebox{\plotpoint}}
\put(334,357){\usebox{\plotpoint}}
\put(335,358){\usebox{\plotpoint}}
\put(336,359){\usebox{\plotpoint}}
\put(337,360){\usebox{\plotpoint}}
\put(338,361){\usebox{\plotpoint}}
\put(339,362){\usebox{\plotpoint}}
\put(340,363){\usebox{\plotpoint}}
\put(341,364){\usebox{\plotpoint}}
\put(342,365){\usebox{\plotpoint}}
\put(343,366){\usebox{\plotpoint}}
\put(344,367){\usebox{\plotpoint}}
\put(345,368){\usebox{\plotpoint}}
\put(346,369){\usebox{\plotpoint}}
\put(347,370){\usebox{\plotpoint}}
\put(348,372){\usebox{\plotpoint}}
\put(349,373){\usebox{\plotpoint}}
\put(350,375){\usebox{\plotpoint}}
\put(351,376){\usebox{\plotpoint}}
\put(352,377){\usebox{\plotpoint}}
\put(353,379){\usebox{\plotpoint}}
\put(354,380){\usebox{\plotpoint}}
\put(355,382){\usebox{\plotpoint}}
\put(356,383){\usebox{\plotpoint}}
\put(357,385){\usebox{\plotpoint}}
\put(358,386){\usebox{\plotpoint}}
\put(359,388){\usebox{\plotpoint}}
\put(360,389){\usebox{\plotpoint}}
\put(361,391){\usebox{\plotpoint}}
\put(362,392){\usebox{\plotpoint}}
\put(363,394){\usebox{\plotpoint}}
\put(364,395){\usebox{\plotpoint}}
\put(365,397){\rule[-0.200pt]{0.400pt}{0.401pt}}
\put(366,398){\rule[-0.200pt]{0.400pt}{0.401pt}}
\put(367,400){\rule[-0.200pt]{0.400pt}{0.401pt}}
\put(368,401){\rule[-0.200pt]{0.400pt}{0.401pt}}
\put(369,403){\rule[-0.200pt]{0.400pt}{0.401pt}}
\put(370,405){\rule[-0.200pt]{0.400pt}{0.401pt}}
\put(371,406){\rule[-0.200pt]{0.400pt}{0.482pt}}
\put(372,409){\rule[-0.200pt]{0.400pt}{0.482pt}}
\put(373,411){\rule[-0.200pt]{0.400pt}{0.482pt}}
\put(374,413){\rule[-0.200pt]{0.400pt}{0.482pt}}
\put(375,415){\rule[-0.200pt]{0.400pt}{0.482pt}}
\put(376,417){\rule[-0.200pt]{0.400pt}{0.401pt}}
\put(377,418){\rule[-0.200pt]{0.400pt}{0.401pt}}
\put(378,420){\rule[-0.200pt]{0.400pt}{0.401pt}}
\put(379,421){\rule[-0.200pt]{0.400pt}{0.401pt}}
\put(380,423){\rule[-0.200pt]{0.400pt}{0.401pt}}
\put(381,425){\rule[-0.200pt]{0.400pt}{0.401pt}}
\put(382,426){\rule[-0.200pt]{0.400pt}{0.442pt}}
\put(383,428){\rule[-0.200pt]{0.400pt}{0.442pt}}
\put(384,430){\rule[-0.200pt]{0.400pt}{0.442pt}}
\put(385,432){\rule[-0.200pt]{0.400pt}{0.442pt}}
\put(386,434){\rule[-0.200pt]{0.400pt}{0.442pt}}
\put(387,436){\rule[-0.200pt]{0.400pt}{0.442pt}}
\put(388,438){\rule[-0.200pt]{0.400pt}{0.401pt}}
\put(389,439){\rule[-0.200pt]{0.400pt}{0.401pt}}
\put(390,441){\rule[-0.200pt]{0.400pt}{0.401pt}}
\put(391,442){\rule[-0.200pt]{0.400pt}{0.401pt}}
\put(392,444){\rule[-0.200pt]{0.400pt}{0.401pt}}
\put(393,446){\rule[-0.200pt]{0.400pt}{0.401pt}}
\put(394,447){\rule[-0.200pt]{0.400pt}{0.442pt}}
\put(395,449){\rule[-0.200pt]{0.400pt}{0.442pt}}
\put(396,451){\rule[-0.200pt]{0.400pt}{0.442pt}}
\put(397,453){\rule[-0.200pt]{0.400pt}{0.442pt}}
\put(398,455){\rule[-0.200pt]{0.400pt}{0.442pt}}
\put(399,457){\rule[-0.200pt]{0.400pt}{0.442pt}}
\put(400,459){\rule[-0.200pt]{0.400pt}{0.401pt}}
\put(401,460){\rule[-0.200pt]{0.400pt}{0.401pt}}
\put(402,462){\rule[-0.200pt]{0.400pt}{0.401pt}}
\put(403,463){\rule[-0.200pt]{0.400pt}{0.401pt}}
\put(404,465){\rule[-0.200pt]{0.400pt}{0.401pt}}
\put(405,467){\rule[-0.200pt]{0.400pt}{0.401pt}}
\put(406,468){\usebox{\plotpoint}}
\put(407,470){\usebox{\plotpoint}}
\put(408,472){\usebox{\plotpoint}}
\put(409,473){\usebox{\plotpoint}}
\put(410,475){\usebox{\plotpoint}}
\put(411,476){\usebox{\plotpoint}}
\put(412,478){\usebox{\plotpoint}}
\put(413,479){\usebox{\plotpoint}}
\put(414,480){\usebox{\plotpoint}}
\put(415,482){\usebox{\plotpoint}}
\put(416,483){\usebox{\plotpoint}}
\put(417,484){\usebox{\plotpoint}}
\put(418,486){\usebox{\plotpoint}}
\put(419,487){\usebox{\plotpoint}}
\put(420,488){\usebox{\plotpoint}}
\put(421,489){\usebox{\plotpoint}}
\put(422,490){\usebox{\plotpoint}}
\put(423,491){\usebox{\plotpoint}}
\put(424,492){\usebox{\plotpoint}}
\put(425,493){\usebox{\plotpoint}}
\put(427,494){\usebox{\plotpoint}}
\put(428,495){\usebox{\plotpoint}}
\put(430,496){\rule[-0.200pt]{1.445pt}{0.400pt}}
\put(436,493){\usebox{\plotpoint}}
\put(437,492){\usebox{\plotpoint}}
\put(438,490){\usebox{\plotpoint}}
\put(439,489){\usebox{\plotpoint}}
\put(440,487){\usebox{\plotpoint}}
\put(441,486){\usebox{\plotpoint}}
\put(442,481){\rule[-0.200pt]{0.400pt}{1.060pt}}
\put(443,477){\rule[-0.200pt]{0.400pt}{1.060pt}}
\put(444,472){\rule[-0.200pt]{0.400pt}{1.060pt}}
\put(445,468){\rule[-0.200pt]{0.400pt}{1.060pt}}
\put(446,464){\rule[-0.200pt]{0.400pt}{1.060pt}}
\put(447,456){\rule[-0.200pt]{0.400pt}{1.887pt}}
\put(448,448){\rule[-0.200pt]{0.400pt}{1.887pt}}
\put(449,440){\rule[-0.200pt]{0.400pt}{1.887pt}}
\put(450,432){\rule[-0.200pt]{0.400pt}{1.887pt}}
\put(451,424){\rule[-0.200pt]{0.400pt}{1.887pt}}
\put(452,417){\rule[-0.200pt]{0.400pt}{1.887pt}}
\put(453,399){\rule[-0.200pt]{0.400pt}{4.216pt}}
\put(454,382){\rule[-0.200pt]{0.400pt}{4.216pt}}
\put(455,364){\rule[-0.200pt]{0.400pt}{4.216pt}}
\put(456,347){\rule[-0.200pt]{0.400pt}{4.216pt}}
\put(457,329){\rule[-0.200pt]{0.400pt}{4.216pt}}
\put(458,312){\rule[-0.200pt]{0.400pt}{4.216pt}}
\put(459,262){\rule[-0.200pt]{0.400pt}{11.985pt}}
\put(460,212){\rule[-0.200pt]{0.400pt}{11.985pt}}
\put(461,162){\rule[-0.200pt]{0.400pt}{11.985pt}}
\put(462,113){\rule[-0.200pt]{0.400pt}{11.985pt}}
\sbox{\plotpoint}{\rule[-0.500pt]{1.000pt}{1.000pt}}%
\put(264,175){\usebox{\plotpoint}}
\put(264,175){\usebox{\plotpoint}}
\put(265,176){\usebox{\plotpoint}}
\put(266,178){\usebox{\plotpoint}}
\put(267,180){\usebox{\plotpoint}}
\put(268,182){\usebox{\plotpoint}}
\put(269,184){\usebox{\plotpoint}}
\put(270,186){\usebox{\plotpoint}}
\put(271,188){\usebox{\plotpoint}}
\put(272,190){\usebox{\plotpoint}}
\put(273,192){\usebox{\plotpoint}}
\put(274,194){\usebox{\plotpoint}}
\put(275,196){\usebox{\plotpoint}}
\put(276,198){\usebox{\plotpoint}}
\put(277,199){\usebox{\plotpoint}}
\put(278,200){\usebox{\plotpoint}}
\put(279,201){\usebox{\plotpoint}}
\put(280,202){\usebox{\plotpoint}}
\put(281,203){\usebox{\plotpoint}}
\put(282,204){\usebox{\plotpoint}}
\put(283,205){\usebox{\plotpoint}}
\put(284,206){\usebox{\plotpoint}}
\put(285,207){\usebox{\plotpoint}}
\put(286,208){\usebox{\plotpoint}}
\put(287,209){\usebox{\plotpoint}}
\put(288,210){\usebox{\plotpoint}}
\put(288,211){\usebox{\plotpoint}}
\put(289,212){\usebox{\plotpoint}}
\put(290,213){\usebox{\plotpoint}}
\put(292,214){\usebox{\plotpoint}}
\put(293,215){\usebox{\plotpoint}}
\put(294,216){\usebox{\plotpoint}}
\put(296,217){\usebox{\plotpoint}}
\put(297,218){\usebox{\plotpoint}}
\put(298,219){\usebox{\plotpoint}}
\put(300,220){\usebox{\plotpoint}}
\put(301,221){\usebox{\plotpoint}}
\put(303,222){\usebox{\plotpoint}}
\put(305,223){\usebox{\plotpoint}}
\put(307,224){\usebox{\plotpoint}}
\put(309,225){\usebox{\plotpoint}}
\put(311,226){\usebox{\plotpoint}}
\put(313,227){\usebox{\plotpoint}}
\put(315,228){\usebox{\plotpoint}}
\put(318,229){\usebox{\plotpoint}}
\put(320,230){\usebox{\plotpoint}}
\put(322,231){\usebox{\plotpoint}}
\put(326,232){\usebox{\plotpoint}}
\put(329,233){\usebox{\plotpoint}}
\put(332,234){\usebox{\plotpoint}}
\put(335,235){\usebox{\plotpoint}}
\put(339,236){\usebox{\plotpoint}}
\put(343,237){\usebox{\plotpoint}}
\put(347,238){\usebox{\plotpoint}}
\put(350,239){\usebox{\plotpoint}}
\put(353,240){\usebox{\plotpoint}}
\put(356,241){\usebox{\plotpoint}}
\put(359,242){\usebox{\plotpoint}}
\put(362,243){\usebox{\plotpoint}}
\put(365,244){\usebox{\plotpoint}}
\put(368,245){\usebox{\plotpoint}}
\put(371,246){\usebox{\plotpoint}}
\put(373,247){\usebox{\plotpoint}}
\put(376,248){\usebox{\plotpoint}}
\put(379,249){\usebox{\plotpoint}}
\put(382,250){\usebox{\plotpoint}}
\put(385,251){\usebox{\plotpoint}}
\put(388,252){\usebox{\plotpoint}}
\put(391,253){\usebox{\plotpoint}}
\put(394,254){\usebox{\plotpoint}}
\put(396,255){\usebox{\plotpoint}}
\put(398,256){\usebox{\plotpoint}}
\put(400,257){\usebox{\plotpoint}}
\put(402,258){\usebox{\plotpoint}}
\put(404,259){\usebox{\plotpoint}}
\put(406,260){\usebox{\plotpoint}}
\put(408,261){\usebox{\plotpoint}}
\put(410,262){\usebox{\plotpoint}}
\put(413,263){\usebox{\plotpoint}}
\put(415,264){\usebox{\plotpoint}}
\put(417,265){\usebox{\plotpoint}}
\put(419,266){\usebox{\plotpoint}}
\put(421,267){\usebox{\plotpoint}}
\put(423,268){\usebox{\plotpoint}}
\put(424,269){\usebox{\plotpoint}}
\put(426,270){\usebox{\plotpoint}}
\put(428,271){\usebox{\plotpoint}}
\put(430,272){\usebox{\plotpoint}}
\put(432,273){\usebox{\plotpoint}}
\put(434,274){\usebox{\plotpoint}}
\put(436,275){\usebox{\plotpoint}}
\put(438,276){\usebox{\plotpoint}}
\put(440,277){\usebox{\plotpoint}}
\put(442,278){\usebox{\plotpoint}}
\put(443,279){\usebox{\plotpoint}}
\put(444,280){\usebox{\plotpoint}}
\put(446,281){\usebox{\plotpoint}}
\put(447,282){\usebox{\plotpoint}}
\put(448,283){\usebox{\plotpoint}}
\put(450,284){\usebox{\plotpoint}}
\put(451,285){\usebox{\plotpoint}}
\put(453,286){\usebox{\plotpoint}}
\put(454,287){\usebox{\plotpoint}}
\put(456,288){\usebox{\plotpoint}}
\put(457,289){\usebox{\plotpoint}}
\put(459,290){\usebox{\plotpoint}}
\put(460,291){\usebox{\plotpoint}}
\put(462,292){\usebox{\plotpoint}}
\put(463,293){\usebox{\plotpoint}}
\put(465,294){\usebox{\plotpoint}}
\put(466,295){\usebox{\plotpoint}}
\put(468,296){\usebox{\plotpoint}}
\put(469,297){\usebox{\plotpoint}}
\put(471,298){\usebox{\plotpoint}}
\put(472,299){\usebox{\plotpoint}}
\put(474,300){\usebox{\plotpoint}}
\put(475,301){\usebox{\plotpoint}}
\put(477,302){\usebox{\plotpoint}}
\put(478,303){\usebox{\plotpoint}}
\put(479,304){\usebox{\plotpoint}}
\put(481,305){\usebox{\plotpoint}}
\put(482,306){\usebox{\plotpoint}}
\put(483,307){\usebox{\plotpoint}}
\put(485,308){\usebox{\plotpoint}}
\put(486,309){\usebox{\plotpoint}}
\put(487,310){\usebox{\plotpoint}}
\put(489,311){\usebox{\plotpoint}}
\put(490,312){\usebox{\plotpoint}}
\put(491,313){\usebox{\plotpoint}}
\put(492,314){\usebox{\plotpoint}}
\put(493,315){\usebox{\plotpoint}}
\put(495,316){\usebox{\plotpoint}}
\put(496,317){\usebox{\plotpoint}}
\put(497,318){\usebox{\plotpoint}}
\put(498,319){\usebox{\plotpoint}}
\put(499,320){\usebox{\plotpoint}}
\put(501,321){\usebox{\plotpoint}}
\put(502,322){\usebox{\plotpoint}}
\put(503,323){\usebox{\plotpoint}}
\put(505,324){\usebox{\plotpoint}}
\put(506,325){\usebox{\plotpoint}}
\put(507,326){\usebox{\plotpoint}}
\put(509,327){\usebox{\plotpoint}}
\put(510,328){\usebox{\plotpoint}}
\put(511,329){\usebox{\plotpoint}}
\put(513,330){\usebox{\plotpoint}}
\put(514,331){\usebox{\plotpoint}}
\put(515,332){\usebox{\plotpoint}}
\put(516,333){\usebox{\plotpoint}}
\put(517,334){\usebox{\plotpoint}}
\put(518,335){\usebox{\plotpoint}}
\put(519,336){\usebox{\plotpoint}}
\put(520,337){\usebox{\plotpoint}}
\put(521,338){\usebox{\plotpoint}}
\put(522,339){\usebox{\plotpoint}}
\put(523,340){\usebox{\plotpoint}}
\put(524,341){\usebox{\plotpoint}}
\put(525,342){\usebox{\plotpoint}}
\put(526,343){\usebox{\plotpoint}}
\put(527,344){\usebox{\plotpoint}}
\put(528,345){\usebox{\plotpoint}}
\put(529,346){\usebox{\plotpoint}}
\put(530,347){\usebox{\plotpoint}}
\put(531,348){\usebox{\plotpoint}}
\put(532,349){\usebox{\plotpoint}}
\put(533,350){\usebox{\plotpoint}}
\put(534,351){\usebox{\plotpoint}}
\put(535,352){\usebox{\plotpoint}}
\put(537,353){\usebox{\plotpoint}}
\put(538,354){\usebox{\plotpoint}}
\put(539,355){\usebox{\plotpoint}}
\put(540,356){\usebox{\plotpoint}}
\put(541,357){\usebox{\plotpoint}}
\put(542,358){\usebox{\plotpoint}}
\put(543,359){\usebox{\plotpoint}}
\put(544,360){\usebox{\plotpoint}}
\put(545,361){\usebox{\plotpoint}}
\put(546,362){\usebox{\plotpoint}}
\put(547,363){\usebox{\plotpoint}}
\put(549,364){\usebox{\plotpoint}}
\put(550,365){\usebox{\plotpoint}}
\put(551,366){\usebox{\plotpoint}}
\put(552,367){\usebox{\plotpoint}}
\put(553,368){\usebox{\plotpoint}}
\put(554,369){\usebox{\plotpoint}}
\put(555,370){\usebox{\plotpoint}}
\put(556,371){\usebox{\plotpoint}}
\put(557,372){\usebox{\plotpoint}}
\put(558,373){\usebox{\plotpoint}}
\put(559,374){\usebox{\plotpoint}}
\put(561,375){\usebox{\plotpoint}}
\put(562,376){\usebox{\plotpoint}}
\put(563,377){\usebox{\plotpoint}}
\put(564,378){\usebox{\plotpoint}}
\put(565,379){\usebox{\plotpoint}}
\put(566,380){\usebox{\plotpoint}}
\put(567,381){\usebox{\plotpoint}}
\put(568,382){\usebox{\plotpoint}}
\put(569,383){\usebox{\plotpoint}}
\put(570,384){\usebox{\plotpoint}}
\put(571,385){\usebox{\plotpoint}}
\put(572,386){\usebox{\plotpoint}}
\put(573,387){\usebox{\plotpoint}}
\put(574,388){\usebox{\plotpoint}}
\put(575,389){\usebox{\plotpoint}}
\put(576,390){\usebox{\plotpoint}}
\put(577,391){\usebox{\plotpoint}}
\put(578,392){\usebox{\plotpoint}}
\put(579,393){\usebox{\plotpoint}}
\put(580,394){\usebox{\plotpoint}}
\put(581,395){\usebox{\plotpoint}}
\put(582,396){\usebox{\plotpoint}}
\put(583,397){\usebox{\plotpoint}}
\put(584,398){\usebox{\plotpoint}}
\put(585,399){\usebox{\plotpoint}}
\put(586,400){\usebox{\plotpoint}}
\put(587,401){\usebox{\plotpoint}}
\put(588,402){\usebox{\plotpoint}}
\put(589,403){\usebox{\plotpoint}}
\put(590,404){\usebox{\plotpoint}}
\put(591,405){\usebox{\plotpoint}}
\put(592,406){\usebox{\plotpoint}}
\put(593,407){\usebox{\plotpoint}}
\put(594,408){\usebox{\plotpoint}}
\put(595,410){\usebox{\plotpoint}}
\put(596,411){\usebox{\plotpoint}}
\put(597,412){\usebox{\plotpoint}}
\put(598,413){\usebox{\plotpoint}}
\put(599,414){\usebox{\plotpoint}}
\put(600,415){\usebox{\plotpoint}}
\put(601,416){\usebox{\plotpoint}}
\put(602,417){\usebox{\plotpoint}}
\put(603,418){\usebox{\plotpoint}}
\put(604,419){\usebox{\plotpoint}}
\put(605,420){\usebox{\plotpoint}}
\put(606,421){\usebox{\plotpoint}}
\put(607,423){\usebox{\plotpoint}}
\put(608,424){\usebox{\plotpoint}}
\put(609,425){\usebox{\plotpoint}}
\put(610,426){\usebox{\plotpoint}}
\put(611,427){\usebox{\plotpoint}}
\put(612,428){\usebox{\plotpoint}}
\put(613,429){\usebox{\plotpoint}}
\put(614,430){\usebox{\plotpoint}}
\put(615,431){\usebox{\plotpoint}}
\put(616,432){\usebox{\plotpoint}}
\put(617,433){\usebox{\plotpoint}}
\put(618,434){\usebox{\plotpoint}}
\put(619,436){\usebox{\plotpoint}}
\put(620,437){\usebox{\plotpoint}}
\put(621,438){\usebox{\plotpoint}}
\put(622,439){\usebox{\plotpoint}}
\put(623,440){\usebox{\plotpoint}}
\put(624,441){\usebox{\plotpoint}}
\put(625,442){\usebox{\plotpoint}}
\put(626,443){\usebox{\plotpoint}}
\put(627,444){\usebox{\plotpoint}}
\put(628,445){\usebox{\plotpoint}}
\put(629,446){\usebox{\plotpoint}}
\put(630,447){\usebox{\plotpoint}}
\put(631,449){\usebox{\plotpoint}}
\put(632,450){\usebox{\plotpoint}}
\put(633,451){\usebox{\plotpoint}}
\put(634,452){\usebox{\plotpoint}}
\put(635,453){\usebox{\plotpoint}}
\put(636,454){\usebox{\plotpoint}}
\put(637,455){\usebox{\plotpoint}}
\put(638,456){\usebox{\plotpoint}}
\put(639,457){\usebox{\plotpoint}}
\put(640,458){\usebox{\plotpoint}}
\put(641,459){\usebox{\plotpoint}}
\put(642,460){\usebox{\plotpoint}}
\put(643,462){\usebox{\plotpoint}}
\put(644,463){\usebox{\plotpoint}}
\put(645,464){\usebox{\plotpoint}}
\put(646,465){\usebox{\plotpoint}}
\put(647,466){\usebox{\plotpoint}}
\put(648,467){\usebox{\plotpoint}}
\put(649,468){\usebox{\plotpoint}}
\put(650,470){\usebox{\plotpoint}}
\put(651,471){\usebox{\plotpoint}}
\put(652,472){\usebox{\plotpoint}}
\put(653,473){\usebox{\plotpoint}}
\put(654,474){\usebox{\plotpoint}}
\put(655,475){\usebox{\plotpoint}}
\put(656,477){\usebox{\plotpoint}}
\put(657,478){\usebox{\plotpoint}}
\put(658,479){\usebox{\plotpoint}}
\put(659,480){\usebox{\plotpoint}}
\put(660,481){\usebox{\plotpoint}}
\put(661,482){\usebox{\plotpoint}}
\put(662,483){\usebox{\plotpoint}}
\put(663,484){\usebox{\plotpoint}}
\put(664,485){\usebox{\plotpoint}}
\put(665,486){\usebox{\plotpoint}}
\put(666,487){\usebox{\plotpoint}}
\put(667,489){\usebox{\plotpoint}}
\put(668,490){\usebox{\plotpoint}}
\put(669,491){\usebox{\plotpoint}}
\put(670,492){\usebox{\plotpoint}}
\put(671,494){\usebox{\plotpoint}}
\put(672,495){\usebox{\plotpoint}}
\put(673,496){\usebox{\plotpoint}}
\put(674,497){\usebox{\plotpoint}}
\put(675,499){\usebox{\plotpoint}}
\put(676,500){\usebox{\plotpoint}}
\put(677,501){\usebox{\plotpoint}}
\put(678,503){\usebox{\plotpoint}}
\put(679,504){\usebox{\plotpoint}}
\put(680,505){\usebox{\plotpoint}}
\put(681,506){\usebox{\plotpoint}}
\put(682,507){\usebox{\plotpoint}}
\put(683,508){\usebox{\plotpoint}}
\put(684,509){\usebox{\plotpoint}}
\put(685,511){\usebox{\plotpoint}}
\put(686,512){\usebox{\plotpoint}}
\put(687,513){\usebox{\plotpoint}}
\put(688,514){\usebox{\plotpoint}}
\put(689,515){\usebox{\plotpoint}}
\put(690,517){\usebox{\plotpoint}}
\put(691,518){\usebox{\plotpoint}}
\put(692,519){\usebox{\plotpoint}}
\put(693,520){\usebox{\plotpoint}}
\put(694,521){\usebox{\plotpoint}}
\put(695,522){\usebox{\plotpoint}}
\put(696,523){\usebox{\plotpoint}}
\put(697,524){\usebox{\plotpoint}}
\put(698,525){\usebox{\plotpoint}}
\put(699,526){\usebox{\plotpoint}}
\put(700,527){\usebox{\plotpoint}}
\put(701,528){\usebox{\plotpoint}}
\put(702,529){\usebox{\plotpoint}}
\put(703,531){\usebox{\plotpoint}}
\put(704,532){\usebox{\plotpoint}}
\put(705,533){\usebox{\plotpoint}}
\put(706,534){\usebox{\plotpoint}}
\put(707,535){\usebox{\plotpoint}}
\put(708,537){\usebox{\plotpoint}}
\put(709,538){\usebox{\plotpoint}}
\put(710,539){\usebox{\plotpoint}}
\put(711,540){\usebox{\plotpoint}}
\put(712,541){\usebox{\plotpoint}}
\put(713,542){\usebox{\plotpoint}}
\put(714,544){\usebox{\plotpoint}}
\put(715,545){\usebox{\plotpoint}}
\put(716,546){\usebox{\plotpoint}}
\put(717,547){\usebox{\plotpoint}}
\put(718,548){\usebox{\plotpoint}}
\put(719,549){\usebox{\plotpoint}}
\put(720,550){\usebox{\plotpoint}}
\put(721,551){\usebox{\plotpoint}}
\put(722,552){\usebox{\plotpoint}}
\put(723,553){\usebox{\plotpoint}}
\put(724,554){\usebox{\plotpoint}}
\put(725,555){\usebox{\plotpoint}}
\put(726,556){\usebox{\plotpoint}}
\put(727,558){\usebox{\plotpoint}}
\put(728,559){\usebox{\plotpoint}}
\put(729,560){\usebox{\plotpoint}}
\put(730,561){\usebox{\plotpoint}}
\put(731,562){\usebox{\plotpoint}}
\put(732,563){\usebox{\plotpoint}}
\put(733,564){\usebox{\plotpoint}}
\put(734,565){\usebox{\plotpoint}}
\put(735,566){\usebox{\plotpoint}}
\put(736,567){\usebox{\plotpoint}}
\put(737,568){\usebox{\plotpoint}}
\put(738,569){\usebox{\plotpoint}}
\put(739,571){\usebox{\plotpoint}}
\put(740,572){\usebox{\plotpoint}}
\put(741,573){\usebox{\plotpoint}}
\put(742,574){\usebox{\plotpoint}}
\put(743,575){\usebox{\plotpoint}}
\put(744,577){\usebox{\plotpoint}}
\put(745,578){\usebox{\plotpoint}}
\put(746,579){\usebox{\plotpoint}}
\put(747,580){\usebox{\plotpoint}}
\put(748,581){\usebox{\plotpoint}}
\put(749,583){\usebox{\plotpoint}}
\put(750,584){\usebox{\plotpoint}}
\put(751,585){\usebox{\plotpoint}}
\put(752,586){\usebox{\plotpoint}}
\put(753,587){\usebox{\plotpoint}}
\put(754,588){\usebox{\plotpoint}}
\put(755,589){\usebox{\plotpoint}}
\put(756,590){\usebox{\plotpoint}}
\put(757,591){\usebox{\plotpoint}}
\put(758,592){\usebox{\plotpoint}}
\put(759,593){\usebox{\plotpoint}}
\put(760,594){\usebox{\plotpoint}}
\put(762,595){\usebox{\plotpoint}}
\put(763,596){\usebox{\plotpoint}}
\put(764,597){\usebox{\plotpoint}}
\put(765,598){\usebox{\plotpoint}}
\put(767,599){\usebox{\plotpoint}}
\put(768,600){\usebox{\plotpoint}}
\put(769,601){\usebox{\plotpoint}}
\put(770,602){\usebox{\plotpoint}}
\put(771,603){\usebox{\plotpoint}}
\put(773,604){\usebox{\plotpoint}}
\put(774,605){\usebox{\plotpoint}}
\put(776,606){\usebox{\plotpoint}}
\put(778,607){\usebox{\plotpoint}}
\put(779,608){\usebox{\plotpoint}}
\put(781,609){\usebox{\plotpoint}}
\put(783,610){\usebox{\plotpoint}}
\put(785,611){\usebox{\plotpoint}}
\put(789,612){\usebox{\plotpoint}}
\put(793,613){\usebox{\plotpoint}}
\put(797,614){\usebox{\plotpoint}}
\put(799,613){\usebox{\plotpoint}}
\put(801,612){\usebox{\plotpoint}}
\put(803,611){\usebox{\plotpoint}}
\put(805,610){\usebox{\plotpoint}}
\put(807,609){\usebox{\plotpoint}}
\put(809,605){\usebox{\plotpoint}}
\put(810,603){\usebox{\plotpoint}}
\put(811,601){\usebox{\plotpoint}}
\put(812,598){\usebox{\plotpoint}}
\put(813,596){\usebox{\plotpoint}}
\put(814,594){\usebox{\plotpoint}}
\put(815,592){\usebox{\plotpoint}}
\put(816,589){\usebox{\plotpoint}}
\put(817,587){\usebox{\plotpoint}}
\put(818,585){\usebox{\plotpoint}}
\put(819,583){\usebox{\plotpoint}}
\put(820,576){\rule[-0.500pt]{1.000pt}{1.566pt}}
\put(821,570){\rule[-0.500pt]{1.000pt}{1.566pt}}
\put(822,563){\rule[-0.500pt]{1.000pt}{1.566pt}}
\put(823,557){\rule[-0.500pt]{1.000pt}{1.566pt}}
\put(824,550){\rule[-0.500pt]{1.000pt}{1.566pt}}
\put(825,544){\rule[-0.500pt]{1.000pt}{1.566pt}}
\put(826,537){\rule[-0.500pt]{1.000pt}{1.566pt}}
\put(827,531){\rule[-0.500pt]{1.000pt}{1.566pt}}
\put(828,524){\rule[-0.500pt]{1.000pt}{1.566pt}}
\put(829,518){\rule[-0.500pt]{1.000pt}{1.566pt}}
\put(830,511){\rule[-0.500pt]{1.000pt}{1.566pt}}
\put(831,505){\rule[-0.500pt]{1.000pt}{1.566pt}}
\put(832,475){\rule[-0.500pt]{1.000pt}{7.086pt}}
\put(833,446){\rule[-0.500pt]{1.000pt}{7.086pt}}
\put(834,416){\rule[-0.500pt]{1.000pt}{7.086pt}}
\put(835,387){\rule[-0.500pt]{1.000pt}{7.086pt}}
\put(836,357){\rule[-0.500pt]{1.000pt}{7.086pt}}
\put(837,328){\rule[-0.500pt]{1.000pt}{7.086pt}}
\put(838,299){\rule[-0.500pt]{1.000pt}{7.086pt}}
\put(839,269){\rule[-0.500pt]{1.000pt}{7.086pt}}
\put(840,240){\rule[-0.500pt]{1.000pt}{7.086pt}}
\put(841,210){\rule[-0.500pt]{1.000pt}{7.086pt}}
\put(842,181){\rule[-0.500pt]{1.000pt}{7.086pt}}
\put(843,152){\rule[-0.500pt]{1.000pt}{7.086pt}}
\put(844,152){\usebox{\plotpoint}}
\sbox{\plotpoint}{\rule[-1.000pt]{2.000pt}{2.000pt}}%
\put(266,113){\usebox{\plotpoint}}
\put(267,114){\usebox{\plotpoint}}
\put(268,116){\usebox{\plotpoint}}
\put(269,118){\usebox{\plotpoint}}
\put(270,120){\usebox{\plotpoint}}
\put(271,122){\usebox{\plotpoint}}
\put(272,123){\usebox{\plotpoint}}
\put(273,125){\usebox{\plotpoint}}
\put(274,127){\usebox{\plotpoint}}
\put(275,129){\usebox{\plotpoint}}
\put(276,131){\usebox{\plotpoint}}
\put(277,132){\usebox{\plotpoint}}
\put(278,133){\usebox{\plotpoint}}
\put(279,134){\usebox{\plotpoint}}
\put(280,136){\usebox{\plotpoint}}
\put(281,137){\usebox{\plotpoint}}
\put(282,138){\usebox{\plotpoint}}
\put(283,139){\usebox{\plotpoint}}
\put(284,141){\usebox{\plotpoint}}
\put(285,142){\usebox{\plotpoint}}
\put(286,143){\usebox{\plotpoint}}
\put(287,144){\usebox{\plotpoint}}
\put(288,146){\usebox{\plotpoint}}
\put(289,147){\usebox{\plotpoint}}
\put(290,148){\usebox{\plotpoint}}
\put(291,149){\usebox{\plotpoint}}
\put(292,150){\usebox{\plotpoint}}
\put(294,151){\usebox{\plotpoint}}
\put(295,152){\usebox{\plotpoint}}
\put(296,153){\usebox{\plotpoint}}
\put(297,154){\usebox{\plotpoint}}
\put(298,155){\usebox{\plotpoint}}
\put(300,156){\usebox{\plotpoint}}
\put(301,157){\usebox{\plotpoint}}
\put(302,158){\usebox{\plotpoint}}
\put(304,159){\usebox{\plotpoint}}
\put(305,160){\usebox{\plotpoint}}
\put(306,161){\usebox{\plotpoint}}
\put(308,162){\usebox{\plotpoint}}
\put(309,163){\usebox{\plotpoint}}
\put(311,164){\usebox{\plotpoint}}
\put(313,165){\usebox{\plotpoint}}
\put(315,166){\usebox{\plotpoint}}
\put(317,167){\usebox{\plotpoint}}
\put(319,168){\usebox{\plotpoint}}
\put(321,169){\usebox{\plotpoint}}
\put(323,170){\usebox{\plotpoint}}
\put(326,171){\usebox{\plotpoint}}
\put(329,172){\usebox{\plotpoint}}
\put(332,173){\usebox{\plotpoint}}
\put(335,174){\usebox{\plotpoint}}
\put(338,175){\usebox{\plotpoint}}
\put(341,176){\usebox{\plotpoint}}
\put(344,177){\usebox{\plotpoint}}
\put(347,178){\usebox{\plotpoint}}
\put(351,179){\usebox{\plotpoint}}
\put(355,180){\usebox{\plotpoint}}
\put(359,181){\usebox{\plotpoint}}
\put(363,182){\usebox{\plotpoint}}
\put(367,183){\usebox{\plotpoint}}
\put(371,184){\usebox{\plotpoint}}
\put(374,185){\usebox{\plotpoint}}
\put(378,186){\usebox{\plotpoint}}
\put(381,187){\usebox{\plotpoint}}
\put(388,188){\usebox{\plotpoint}}
\put(394,189){\usebox{\plotpoint}}
\put(400,190){\usebox{\plotpoint}}
\put(406,191){\usebox{\plotpoint}}
\put(410,192){\usebox{\plotpoint}}
\put(414,193){\usebox{\plotpoint}}
\put(418,194){\usebox{\plotpoint}}
\put(424,195){\usebox{\plotpoint}}
\put(430,196){\usebox{\plotpoint}}
\put(436,197){\usebox{\plotpoint}}
\put(442,198){\usebox{\plotpoint}}
\put(445,199){\usebox{\plotpoint}}
\put(449,200){\usebox{\plotpoint}}
\put(452,201){\usebox{\plotpoint}}
\put(457,202){\usebox{\plotpoint}}
\put(461,203){\usebox{\plotpoint}}
\put(465,204){\usebox{\plotpoint}}
\put(471,205){\usebox{\plotpoint}}
\put(477,206){\usebox{\plotpoint}}
\put(481,207){\usebox{\plotpoint}}
\put(485,208){\usebox{\plotpoint}}
\put(489,209){\usebox{\plotpoint}}
\put(493,210){\usebox{\plotpoint}}
\put(497,211){\usebox{\plotpoint}}
\put(501,212){\usebox{\plotpoint}}
\put(504,213){\usebox{\plotpoint}}
\put(507,214){\usebox{\plotpoint}}
\put(510,215){\usebox{\plotpoint}}
\put(513,216){\usebox{\plotpoint}}
\put(516,217){\usebox{\plotpoint}}
\put(520,218){\usebox{\plotpoint}}
\put(524,219){\usebox{\plotpoint}}
\put(528,220){\usebox{\plotpoint}}
\put(532,221){\usebox{\plotpoint}}
\put(536,222){\usebox{\plotpoint}}
\put(539,223){\usebox{\plotpoint}}
\put(542,224){\usebox{\plotpoint}}
\put(545,225){\usebox{\plotpoint}}
\put(548,226){\usebox{\plotpoint}}
\put(551,227){\usebox{\plotpoint}}
\put(554,228){\usebox{\plotpoint}}
\put(557,229){\usebox{\plotpoint}}
\put(560,230){\usebox{\plotpoint}}
\put(563,231){\usebox{\plotpoint}}
\put(566,232){\usebox{\plotpoint}}
\put(569,233){\usebox{\plotpoint}}
\put(572,234){\usebox{\plotpoint}}
\put(575,235){\usebox{\plotpoint}}
\put(578,236){\usebox{\plotpoint}}
\put(581,237){\usebox{\plotpoint}}
\put(584,238){\usebox{\plotpoint}}
\put(586,239){\usebox{\plotpoint}}
\put(589,240){\usebox{\plotpoint}}
\put(592,241){\usebox{\plotpoint}}
\put(595,242){\usebox{\plotpoint}}
\put(597,243){\usebox{\plotpoint}}
\put(599,244){\usebox{\plotpoint}}
\put(602,245){\usebox{\plotpoint}}
\put(604,246){\usebox{\plotpoint}}
\put(607,247){\usebox{\plotpoint}}
\put(610,248){\usebox{\plotpoint}}
\put(613,249){\usebox{\plotpoint}}
\put(616,250){\usebox{\plotpoint}}
\put(619,251){\usebox{\plotpoint}}
\put(621,252){\usebox{\plotpoint}}
\put(623,253){\usebox{\plotpoint}}
\put(626,254){\usebox{\plotpoint}}
\put(628,255){\usebox{\plotpoint}}
\put(631,256){\usebox{\plotpoint}}
\put(633,257){\usebox{\plotpoint}}
\put(635,258){\usebox{\plotpoint}}
\put(638,259){\usebox{\plotpoint}}
\put(640,260){\usebox{\plotpoint}}
\put(643,261){\usebox{\plotpoint}}
\put(645,262){\usebox{\plotpoint}}
\put(647,263){\usebox{\plotpoint}}
\put(650,264){\usebox{\plotpoint}}
\put(652,265){\usebox{\plotpoint}}
\put(655,266){\usebox{\plotpoint}}
\put(657,267){\usebox{\plotpoint}}
\put(659,268){\usebox{\plotpoint}}
\put(662,269){\usebox{\plotpoint}}
\put(664,270){\usebox{\plotpoint}}
\put(667,271){\usebox{\plotpoint}}
\put(669,272){\usebox{\plotpoint}}
\put(671,273){\usebox{\plotpoint}}
\put(673,274){\usebox{\plotpoint}}
\put(675,275){\usebox{\plotpoint}}
\put(678,276){\usebox{\plotpoint}}
\put(680,277){\usebox{\plotpoint}}
\put(682,278){\usebox{\plotpoint}}
\put(684,279){\usebox{\plotpoint}}
\put(686,280){\usebox{\plotpoint}}
\put(688,281){\usebox{\plotpoint}}
\put(690,282){\usebox{\plotpoint}}
\put(692,283){\usebox{\plotpoint}}
\put(694,284){\usebox{\plotpoint}}
\put(697,285){\usebox{\plotpoint}}
\put(699,286){\usebox{\plotpoint}}
\put(702,287){\usebox{\plotpoint}}
\put(704,288){\usebox{\plotpoint}}
\put(706,289){\usebox{\plotpoint}}
\put(708,290){\usebox{\plotpoint}}
\put(710,291){\usebox{\plotpoint}}
\put(712,292){\usebox{\plotpoint}}
\put(714,293){\usebox{\plotpoint}}
\put(716,294){\usebox{\plotpoint}}
\put(718,295){\usebox{\plotpoint}}
\put(720,296){\usebox{\plotpoint}}
\put(722,297){\usebox{\plotpoint}}
\put(724,298){\usebox{\plotpoint}}
\put(726,299){\usebox{\plotpoint}}
\put(728,300){\usebox{\plotpoint}}
\put(730,301){\usebox{\plotpoint}}
\put(732,302){\usebox{\plotpoint}}
\put(734,303){\usebox{\plotpoint}}
\put(736,304){\usebox{\plotpoint}}
\put(738,305){\usebox{\plotpoint}}
\put(739,306){\usebox{\plotpoint}}
\put(741,307){\usebox{\plotpoint}}
\put(743,308){\usebox{\plotpoint}}
\put(745,309){\usebox{\plotpoint}}
\put(747,310){\usebox{\plotpoint}}
\put(748,311){\usebox{\plotpoint}}
\put(751,312){\usebox{\plotpoint}}
\put(753,313){\usebox{\plotpoint}}
\put(755,314){\usebox{\plotpoint}}
\put(757,315){\usebox{\plotpoint}}
\put(759,316){\usebox{\plotpoint}}
\put(761,317){\usebox{\plotpoint}}
\put(763,318){\usebox{\plotpoint}}
\put(765,319){\usebox{\plotpoint}}
\put(767,320){\usebox{\plotpoint}}
\put(769,321){\usebox{\plotpoint}}
\put(771,322){\usebox{\plotpoint}}
\put(773,323){\usebox{\plotpoint}}
\put(774,324){\usebox{\plotpoint}}
\put(776,325){\usebox{\plotpoint}}
\put(778,326){\usebox{\plotpoint}}
\put(779,327){\usebox{\plotpoint}}
\put(781,328){\usebox{\plotpoint}}
\put(783,329){\usebox{\plotpoint}}
\put(785,330){\usebox{\plotpoint}}
\put(787,331){\usebox{\plotpoint}}
\put(789,332){\usebox{\plotpoint}}
\put(791,333){\usebox{\plotpoint}}
\put(793,334){\usebox{\plotpoint}}
\put(795,335){\usebox{\plotpoint}}
\put(797,336){\usebox{\plotpoint}}
\put(798,337){\usebox{\plotpoint}}
\put(800,338){\usebox{\plotpoint}}
\put(802,339){\usebox{\plotpoint}}
\put(803,340){\usebox{\plotpoint}}
\put(805,341){\usebox{\plotpoint}}
\put(807,342){\usebox{\plotpoint}}
\put(809,343){\usebox{\plotpoint}}
\put(810,344){\usebox{\plotpoint}}
\put(812,345){\usebox{\plotpoint}}
\put(813,346){\usebox{\plotpoint}}
\put(815,347){\usebox{\plotpoint}}
\put(816,348){\usebox{\plotpoint}}
\put(818,349){\usebox{\plotpoint}}
\put(819,350){\usebox{\plotpoint}}
\put(822,351){\usebox{\plotpoint}}
\put(824,352){\usebox{\plotpoint}}
\put(826,353){\usebox{\plotpoint}}
\put(828,354){\usebox{\plotpoint}}
\put(830,355){\usebox{\plotpoint}}
\put(832,356){\usebox{\plotpoint}}
\put(833,357){\usebox{\plotpoint}}
\put(835,358){\usebox{\plotpoint}}
\put(837,359){\usebox{\plotpoint}}
\put(838,360){\usebox{\plotpoint}}
\put(840,361){\usebox{\plotpoint}}
\put(842,362){\usebox{\plotpoint}}
\put(844,363){\usebox{\plotpoint}}
\put(845,364){\usebox{\plotpoint}}
\put(847,365){\usebox{\plotpoint}}
\put(849,366){\usebox{\plotpoint}}
\put(850,367){\usebox{\plotpoint}}
\put(852,368){\usebox{\plotpoint}}
\put(854,369){\usebox{\plotpoint}}
\put(856,370){\usebox{\plotpoint}}
\put(857,371){\usebox{\plotpoint}}
\put(859,372){\usebox{\plotpoint}}
\put(861,373){\usebox{\plotpoint}}
\put(862,374){\usebox{\plotpoint}}
\put(864,375){\usebox{\plotpoint}}
\put(866,376){\usebox{\plotpoint}}
\put(868,377){\usebox{\plotpoint}}
\put(869,378){\usebox{\plotpoint}}
\put(871,379){\usebox{\plotpoint}}
\put(873,380){\usebox{\plotpoint}}
\put(874,381){\usebox{\plotpoint}}
\put(876,382){\usebox{\plotpoint}}
\put(878,383){\usebox{\plotpoint}}
\put(880,384){\usebox{\plotpoint}}
\put(881,385){\usebox{\plotpoint}}
\put(883,386){\usebox{\plotpoint}}
\put(884,387){\usebox{\plotpoint}}
\put(886,388){\usebox{\plotpoint}}
\put(887,389){\usebox{\plotpoint}}
\put(889,390){\usebox{\plotpoint}}
\put(890,391){\usebox{\plotpoint}}
\put(892,392){\usebox{\plotpoint}}
\put(894,393){\usebox{\plotpoint}}
\put(895,394){\usebox{\plotpoint}}
\put(897,395){\usebox{\plotpoint}}
\put(898,396){\usebox{\plotpoint}}
\put(900,397){\usebox{\plotpoint}}
\put(901,398){\usebox{\plotpoint}}
\put(903,399){\usebox{\plotpoint}}
\put(904,400){\usebox{\plotpoint}}
\put(906,401){\usebox{\plotpoint}}
\put(908,402){\usebox{\plotpoint}}
\put(909,403){\usebox{\plotpoint}}
\put(911,404){\usebox{\plotpoint}}
\put(913,405){\usebox{\plotpoint}}
\put(915,406){\usebox{\plotpoint}}
\put(916,407){\usebox{\plotpoint}}
\put(918,408){\usebox{\plotpoint}}
\put(920,409){\usebox{\plotpoint}}
\put(921,410){\usebox{\plotpoint}}
\put(923,411){\usebox{\plotpoint}}
\put(925,412){\usebox{\plotpoint}}
\put(927,413){\usebox{\plotpoint}}
\put(928,414){\usebox{\plotpoint}}
\put(930,415){\usebox{\plotpoint}}
\put(931,416){\usebox{\plotpoint}}
\put(933,417){\usebox{\plotpoint}}
\put(934,418){\usebox{\plotpoint}}
\put(936,419){\usebox{\plotpoint}}
\put(937,420){\usebox{\plotpoint}}
\put(939,421){\usebox{\plotpoint}}
\put(940,422){\usebox{\plotpoint}}
\put(942,423){\usebox{\plotpoint}}
\put(944,424){\usebox{\plotpoint}}
\put(945,425){\usebox{\plotpoint}}
\put(947,426){\usebox{\plotpoint}}
\put(949,427){\usebox{\plotpoint}}
\put(951,428){\usebox{\plotpoint}}
\put(952,429){\usebox{\plotpoint}}
\put(953,430){\usebox{\plotpoint}}
\put(955,431){\usebox{\plotpoint}}
\put(956,432){\usebox{\plotpoint}}
\put(957,433){\usebox{\plotpoint}}
\put(959,434){\usebox{\plotpoint}}
\put(960,435){\usebox{\plotpoint}}
\put(962,436){\usebox{\plotpoint}}
\put(963,437){\usebox{\plotpoint}}
\put(965,438){\usebox{\plotpoint}}
\put(966,439){\usebox{\plotpoint}}
\put(968,440){\usebox{\plotpoint}}
\put(969,441){\usebox{\plotpoint}}
\put(971,442){\usebox{\plotpoint}}
\put(972,443){\usebox{\plotpoint}}
\put(974,444){\usebox{\plotpoint}}
\put(975,445){\usebox{\plotpoint}}
\put(977,446){\usebox{\plotpoint}}
\put(979,447){\usebox{\plotpoint}}
\put(980,448){\usebox{\plotpoint}}
\put(982,449){\usebox{\plotpoint}}
\put(984,450){\usebox{\plotpoint}}
\put(986,451){\usebox{\plotpoint}}
\put(987,452){\usebox{\plotpoint}}
\put(989,453){\usebox{\plotpoint}}
\put(990,454){\usebox{\plotpoint}}
\put(992,455){\usebox{\plotpoint}}
\put(993,456){\usebox{\plotpoint}}
\put(995,457){\usebox{\plotpoint}}
\put(996,458){\usebox{\plotpoint}}
\put(998,459){\usebox{\plotpoint}}
\put(999,460){\usebox{\plotpoint}}
\put(1001,461){\usebox{\plotpoint}}
\put(1002,462){\usebox{\plotpoint}}
\put(1004,463){\usebox{\plotpoint}}
\put(1005,464){\usebox{\plotpoint}}
\put(1007,465){\usebox{\plotpoint}}
\put(1008,466){\usebox{\plotpoint}}
\put(1010,467){\usebox{\plotpoint}}
\put(1011,468){\usebox{\plotpoint}}
\put(1013,469){\usebox{\plotpoint}}
\put(1014,470){\usebox{\plotpoint}}
\put(1016,471){\usebox{\plotpoint}}
\put(1017,472){\usebox{\plotpoint}}
\put(1019,473){\usebox{\plotpoint}}
\put(1020,474){\usebox{\plotpoint}}
\put(1022,475){\usebox{\plotpoint}}
\put(1023,476){\usebox{\plotpoint}}
\put(1024,477){\usebox{\plotpoint}}
\put(1026,478){\usebox{\plotpoint}}
\put(1027,479){\usebox{\plotpoint}}
\put(1028,480){\usebox{\plotpoint}}
\put(1030,481){\usebox{\plotpoint}}
\put(1031,482){\usebox{\plotpoint}}
\put(1033,483){\usebox{\plotpoint}}
\put(1034,484){\usebox{\plotpoint}}
\put(1036,485){\usebox{\plotpoint}}
\put(1037,486){\usebox{\plotpoint}}
\put(1039,487){\usebox{\plotpoint}}
\put(1040,488){\usebox{\plotpoint}}
\put(1042,489){\usebox{\plotpoint}}
\put(1043,490){\usebox{\plotpoint}}
\put(1045,491){\usebox{\plotpoint}}
\put(1046,492){\usebox{\plotpoint}}
\put(1048,493){\usebox{\plotpoint}}
\put(1049,494){\usebox{\plotpoint}}
\put(1051,495){\usebox{\plotpoint}}
\put(1052,496){\usebox{\plotpoint}}
\put(1054,497){\usebox{\plotpoint}}
\put(1055,498){\usebox{\plotpoint}}
\put(1057,499){\usebox{\plotpoint}}
\put(1058,500){\usebox{\plotpoint}}
\put(1060,501){\usebox{\plotpoint}}
\put(1061,502){\usebox{\plotpoint}}
\put(1063,503){\usebox{\plotpoint}}
\put(1064,504){\usebox{\plotpoint}}
\put(1066,505){\usebox{\plotpoint}}
\put(1067,506){\usebox{\plotpoint}}
\put(1069,507){\usebox{\plotpoint}}
\put(1070,508){\usebox{\plotpoint}}
\put(1072,509){\usebox{\plotpoint}}
\put(1073,510){\usebox{\plotpoint}}
\put(1075,511){\usebox{\plotpoint}}
\put(1076,512){\usebox{\plotpoint}}
\put(1078,513){\usebox{\plotpoint}}
\put(1079,514){\usebox{\plotpoint}}
\put(1081,515){\usebox{\plotpoint}}
\put(1082,516){\usebox{\plotpoint}}
\put(1083,517){\usebox{\plotpoint}}
\put(1085,518){\usebox{\plotpoint}}
\put(1086,519){\usebox{\plotpoint}}
\put(1087,520){\usebox{\plotpoint}}
\put(1089,521){\usebox{\plotpoint}}
\put(1090,522){\usebox{\plotpoint}}
\put(1091,523){\usebox{\plotpoint}}
\put(1093,524){\usebox{\plotpoint}}
\put(1094,525){\usebox{\plotpoint}}
\put(1096,526){\usebox{\plotpoint}}
\put(1097,527){\usebox{\plotpoint}}
\put(1099,528){\usebox{\plotpoint}}
\put(1100,529){\usebox{\plotpoint}}
\put(1102,530){\usebox{\plotpoint}}
\put(1103,531){\usebox{\plotpoint}}
\put(1105,532){\usebox{\plotpoint}}
\put(1106,533){\usebox{\plotpoint}}
\put(1107,534){\usebox{\plotpoint}}
\put(1109,535){\usebox{\plotpoint}}
\put(1110,536){\usebox{\plotpoint}}
\put(1111,537){\usebox{\plotpoint}}
\put(1113,538){\usebox{\plotpoint}}
\put(1114,539){\usebox{\plotpoint}}
\put(1116,540){\usebox{\plotpoint}}
\put(1117,541){\usebox{\plotpoint}}
\put(1119,542){\usebox{\plotpoint}}
\put(1120,543){\usebox{\plotpoint}}
\put(1122,544){\usebox{\plotpoint}}
\put(1123,545){\usebox{\plotpoint}}
\put(1125,546){\usebox{\plotpoint}}
\put(1126,547){\usebox{\plotpoint}}
\put(1128,548){\usebox{\plotpoint}}
\put(1129,549){\usebox{\plotpoint}}
\put(1130,550){\usebox{\plotpoint}}
\put(1132,551){\usebox{\plotpoint}}
\put(1133,552){\usebox{\plotpoint}}
\put(1134,553){\usebox{\plotpoint}}
\put(1136,554){\usebox{\plotpoint}}
\put(1137,555){\usebox{\plotpoint}}
\put(1138,556){\usebox{\plotpoint}}
\put(1140,557){\usebox{\plotpoint}}
\put(1141,558){\usebox{\plotpoint}}
\put(1143,559){\usebox{\plotpoint}}
\put(1144,560){\usebox{\plotpoint}}
\put(1146,561){\usebox{\plotpoint}}
\put(1147,562){\usebox{\plotpoint}}
\put(1149,563){\usebox{\plotpoint}}
\put(1150,564){\usebox{\plotpoint}}
\put(1152,565){\usebox{\plotpoint}}
\put(1153,566){\usebox{\plotpoint}}
\put(1155,567){\usebox{\plotpoint}}
\put(1156,568){\usebox{\plotpoint}}
\put(1158,569){\usebox{\plotpoint}}
\put(1159,570){\usebox{\plotpoint}}
\put(1161,571){\usebox{\plotpoint}}
\put(1162,572){\usebox{\plotpoint}}
\put(1164,573){\usebox{\plotpoint}}
\put(1165,574){\usebox{\plotpoint}}
\put(1166,575){\usebox{\plotpoint}}
\put(1168,576){\usebox{\plotpoint}}
\put(1169,577){\usebox{\plotpoint}}
\put(1170,578){\usebox{\plotpoint}}
\put(1172,579){\usebox{\plotpoint}}
\put(1173,580){\usebox{\plotpoint}}
\put(1174,581){\usebox{\plotpoint}}
\put(1176,582){\usebox{\plotpoint}}
\put(1177,583){\usebox{\plotpoint}}
\put(1178,584){\usebox{\plotpoint}}
\put(1180,585){\usebox{\plotpoint}}
\put(1181,586){\usebox{\plotpoint}}
\put(1182,587){\usebox{\plotpoint}}
\put(1184,588){\usebox{\plotpoint}}
\put(1185,589){\usebox{\plotpoint}}
\put(1187,590){\usebox{\plotpoint}}
\put(1188,591){\usebox{\plotpoint}}
\put(1190,592){\usebox{\plotpoint}}
\put(1191,593){\usebox{\plotpoint}}
\put(1193,594){\usebox{\plotpoint}}
\put(1194,595){\usebox{\plotpoint}}
\put(1196,596){\usebox{\plotpoint}}
\put(1197,597){\usebox{\plotpoint}}
\put(1199,598){\usebox{\plotpoint}}
\put(1200,599){\usebox{\plotpoint}}
\put(1202,600){\usebox{\plotpoint}}
\put(1203,601){\usebox{\plotpoint}}
\put(1205,602){\usebox{\plotpoint}}
\put(1206,603){\usebox{\plotpoint}}
\put(1208,604){\usebox{\plotpoint}}
\put(1209,605){\usebox{\plotpoint}}
\put(1211,606){\usebox{\plotpoint}}
\put(1212,607){\usebox{\plotpoint}}
\put(1214,608){\usebox{\plotpoint}}
\put(1215,609){\usebox{\plotpoint}}
\put(1217,610){\usebox{\plotpoint}}
\put(1218,611){\usebox{\plotpoint}}
\put(1220,612){\usebox{\plotpoint}}
\put(1221,613){\usebox{\plotpoint}}
\put(1223,614){\usebox{\plotpoint}}
\put(1224,615){\usebox{\plotpoint}}
\put(1226,616){\usebox{\plotpoint}}
\put(1227,617){\usebox{\plotpoint}}
\put(1229,618){\usebox{\plotpoint}}
\put(1230,619){\usebox{\plotpoint}}
\put(1232,620){\usebox{\plotpoint}}
\put(1233,621){\usebox{\plotpoint}}
\put(1235,622){\usebox{\plotpoint}}
\put(1236,623){\usebox{\plotpoint}}
\put(1238,624){\usebox{\plotpoint}}
\put(1239,625){\usebox{\plotpoint}}
\put(1241,626){\usebox{\plotpoint}}
\put(1242,627){\usebox{\plotpoint}}
\put(1244,628){\usebox{\plotpoint}}
\put(1245,629){\usebox{\plotpoint}}
\put(1247,630){\usebox{\plotpoint}}
\put(1248,631){\usebox{\plotpoint}}
\put(1250,632){\usebox{\plotpoint}}
\put(1251,633){\usebox{\plotpoint}}
\put(1253,634){\usebox{\plotpoint}}
\put(1254,635){\usebox{\plotpoint}}
\put(1256,636){\usebox{\plotpoint}}
\put(1257,637){\usebox{\plotpoint}}
\put(1259,638){\usebox{\plotpoint}}
\put(1261,639){\usebox{\plotpoint}}
\put(1263,640){\usebox{\plotpoint}}
\put(1264,641){\usebox{\plotpoint}}
\put(1266,642){\usebox{\plotpoint}}
\put(1268,643){\usebox{\plotpoint}}
\put(1269,644){\usebox{\plotpoint}}
\put(1272,645){\usebox{\plotpoint}}
\put(1274,646){\usebox{\plotpoint}}
\put(1276,647){\usebox{\plotpoint}}
\put(1278,648){\usebox{\plotpoint}}
\put(1280,649){\usebox{\plotpoint}}
\put(1282,650){\usebox{\plotpoint}}
\put(1284,651){\usebox{\plotpoint}}
\put(1286,652){\usebox{\plotpoint}}
\put(1289,653){\usebox{\plotpoint}}
\put(1291,654){\usebox{\plotpoint}}
\put(1294,655){\usebox{\plotpoint}}
\put(1297,656){\usebox{\plotpoint}}
\put(1300,657){\usebox{\plotpoint}}
\put(1303,658){\usebox{\plotpoint}}
\put(1306,659){\rule[-1.000pt]{2.891pt}{2.000pt}}
\put(1318,660){\rule[-1.000pt]{2.650pt}{2.000pt}}
\put(1329,659){\usebox{\plotpoint}}
\put(1330,658){\usebox{\plotpoint}}
\put(1332,657){\usebox{\plotpoint}}
\put(1333,656){\usebox{\plotpoint}}
\put(1335,655){\usebox{\plotpoint}}
\put(1336,654){\usebox{\plotpoint}}
\put(1338,653){\usebox{\plotpoint}}
\put(1339,652){\usebox{\plotpoint}}
\put(1341,649){\usebox{\plotpoint}}
\put(1342,648){\usebox{\plotpoint}}
\put(1343,646){\usebox{\plotpoint}}
\put(1344,645){\usebox{\plotpoint}}
\put(1345,643){\usebox{\plotpoint}}
\put(1346,642){\usebox{\plotpoint}}
\put(1347,640){\usebox{\plotpoint}}
\put(1348,639){\usebox{\plotpoint}}
\put(1349,637){\usebox{\plotpoint}}
\put(1350,636){\usebox{\plotpoint}}
\put(1351,634){\usebox{\plotpoint}}
\put(1352,633){\usebox{\plotpoint}}
\put(1353,629){\usebox{\plotpoint}}
\put(1354,625){\usebox{\plotpoint}}
\put(1355,622){\usebox{\plotpoint}}
\put(1356,618){\usebox{\plotpoint}}
\put(1357,615){\usebox{\plotpoint}}
\put(1358,611){\usebox{\plotpoint}}
\put(1359,607){\usebox{\plotpoint}}
\put(1360,604){\usebox{\plotpoint}}
\put(1361,600){\usebox{\plotpoint}}
\put(1362,597){\usebox{\plotpoint}}
\put(1363,593){\usebox{\plotpoint}}
\put(1364,590){\usebox{\plotpoint}}
\put(1365,580){\rule[-1.000pt]{2.000pt}{2.389pt}}
\put(1366,570){\rule[-1.000pt]{2.000pt}{2.389pt}}
\put(1367,560){\rule[-1.000pt]{2.000pt}{2.389pt}}
\put(1368,550){\rule[-1.000pt]{2.000pt}{2.389pt}}
\put(1369,540){\rule[-1.000pt]{2.000pt}{2.389pt}}
\put(1370,530){\rule[-1.000pt]{2.000pt}{2.389pt}}
\put(1371,520){\rule[-1.000pt]{2.000pt}{2.389pt}}
\put(1372,510){\rule[-1.000pt]{2.000pt}{2.389pt}}
\put(1373,500){\rule[-1.000pt]{2.000pt}{2.389pt}}
\put(1374,490){\rule[-1.000pt]{2.000pt}{2.389pt}}
\put(1375,480){\rule[-1.000pt]{2.000pt}{2.389pt}}
\put(1376,471){\rule[-1.000pt]{2.000pt}{2.389pt}}
\put(1377,411){\rule[-1.000pt]{2.000pt}{14.374pt}}
\put(1378,351){\rule[-1.000pt]{2.000pt}{14.374pt}}
\put(1379,292){\rule[-1.000pt]{2.000pt}{14.374pt}}
\put(1380,232){\rule[-1.000pt]{2.000pt}{14.374pt}}
\put(1381,172){\rule[-1.000pt]{2.000pt}{14.374pt}}
\put(1382,113){\rule[-1.000pt]{2.000pt}{14.374pt}}
\put(1383,113){\usebox{\plotpoint}}
\end{picture}